\documentclass[aps,prc]{revtex4}

\def\blfootnote{\xdef\@thefnmark{}\@footnotetext}

\long\def\symbolfootnote[#1]#2{\begingroup%
\def\thefootnote{\fnsymbol{footnote}}\footnote[#1]{#2}\endgroup}

\def\lsim{ \,\, \vcenter{\hbox{$\buildrel{\displaystyle <}\over\sim$}}
 \,\,}
\def\gsim{ \,\, \vcenter{\hbox{$\buildrel{\displaystyle >}\over\sim$}}
 \,\,}
\newcommand{\be}{\begin{eqnarray}}
\newcommand{\ee}{\end{eqnarray}}
\newcommand{\bcent}{\begin{center}}
\newcommand{\ecent}{\end{center}}
\newcommand{\benum}{\begin{enumerate}}
\newcommand{\eenum}{\end{enumerate}}
\newcommand{\bdesc}{\begin{description}}
\newcommand{\edesc}{\end{description}}
\newcommand{\bitem}{\begin{itemize}}
\newcommand{\eitem}{\end{itemize}}
\newcommand{\bhead}{\begin{center}\bf \Large}
\newcommand{\ehead}{\end{center}\bigskip}
\newcommand{\non}{\nonumber\\}

 \newcommand{\bfk}{{\bf k}}

 \newcommand{\bfn}{{\bf n}} 
  
 \newcommand{\bfp}{{\bf p}} 
  
 \newcommand{\bfr}{{\bf r}}

 \newcommand{\calP}{{\cal P}}

 \newcommand{\eqn}[1]{Eq.~(\ref{#1})}

 \newcommand{\ave}[1]{\langle {#1} \rangle}
 \newcommand{\MeV}{\hbox{MeV}}
 \newcommand{\GeV}{\hbox{GeV}}
 
 \newcommand{\fm}{\hbox{fm}}

 \newcommand{\tw}{\textwidth}
 
\newcommand{\gs}{g_{\rm s}}
\newcommand{\mD}{m_{\rm D}}
\newcommand{\q}{{\bf q}}
\newcommand{\p}{{\bf p}}
\newcommand{\F}{{\bf F}}
\newcommand{\h}{{\bf h}}
\def\Pq{P_{\!\!q \bar q}}
\def\Pg{P_{\! g}}

\usepackage{epsfig}
\usepackage{psfig}

\begin{document}
\pagestyle{empty}
 \title{Energy Loss of Leading Hadrons and Direct Photon production
 in Evolving Quark-Gluon Plasma}
\author
{Simon Turbide$^1$, Charles Gale$^{1}$, Sangyong Jeon$^{1,2}$ and Guy D.
Moore$^1$}
\affiliation
{$^1$ Department of Physics, McGill University, 3600 University Street,
    Montreal, Canada H3A 2T8\\
 $^2$ RIKEN-BNL Research Center, Upton, NY 11973-5000}

\date{\today}

\begin{abstract}
We calculate the nuclear modification factor of neutral pions and the
photon yield at high $p_T$ in central
Au-Au collisions at RHIC ($\sqrt{s}=$200 GeV) and Pb-Pb collisions at
the LHC
($\sqrt{s}=$5500 GeV).  A leading-order accurate treatment of jet energy loss
in the medium has been convolved with a physical description of the initial
spatial distribution of jets and a (1+1) dimensional expansion.  We
reproduce the
nuclear modification factor of pion $R_{AA}$ at RHIC, assuming an initial
temperature $T_i=$370 MeV and a formation time $\tau_i=$0.26 fm/c,
corresponding to $dN/dy=1260$.  The resulting suppression depends on the
particle rapidity density $dN/dy$ but weakly on the initial temperature.  The
jet energy loss treatment is also included in the calculation of high $p_T$
photons.  Photons coming from primordial hard N-N scattering are the dominant
contribution at RHIC for $p_T > $ 5 GeV, while at the LHC, the range
$8<p_T<14$ GeV is dominated by jet-photon conversion in the plasma.

\end{abstract}

 \maketitle

%\pacs{}

\pagestyle{plain}

\section{Introduction}
\setcounter{page}{1}

 One of the most important issues that arises in the study of 
 relativistic heavy ion collisions is
 that of 
 the creation of a quark-gluon plasma (QGP).
 From lattice QCD calculations, there is strong evidence that at
 around a temperature
 of $170\,\MeV$, or equivalently an energy density of about $1\,\GeV/\fm^3$
{}~\cite{karsch},
 there should be
 a phase transition between an ordinary confined hadronic matter phase 
%the confined (ordinary matter) phase
 and a QGP phase, with an associated change in the relevant degrees of freedom.
 At the Relativistic Heavy Ion Collider (RHIC), measurements support the
 assertion that the initial energy density of the {\em created} system
 reaches up to
 $5\,\GeV/\fm^3$.  However, this does not necessarily
 imply the creation of a QGP.
 To make sure that a new state of matter has been formed, independent evidence
 is needed. 
 One of the most striking measurements in support of the creation of hot and
 dense matter at RHIC 
 is the discovery of high $p_T$ suppression in
 central Au-Au collisions.  This phenomenon is observed in single hadron
 spectra~\cite{phenix2,star} and in the disappearance of back-to-back
 correlations of
 high $p_T$ hadrons~\cite{star2}.

 There are a number of factors that can potentially influence the spectrum of
 high $p_T$ partons in heavy ion collisions compared to that in
 hadron-hadron collisions:
 \bitem
  \item[(i)] A difference between the parton distribution functions of a proton
  and a heavy nucleus.  This can be both depletion (shadowing) or excess
  (anti-shadowing) depending on the value of momentum fraction $x$.
  This also includes the effect of gluon saturation.

  \item[(ii)] The initial state multiple scattering effect. This is the
  well known Cronin effect caused by multiple soft scatterings a parton
  may suffer before it makes a hard collision. This can, of course, only
  be significant in interactions involving nuclei.

 \item[(iii)] Final state energy loss. This is due to the interaction
 between the produced hard parton and the hot and dense environment.
 \eitem
\noindent

Our kinematical range of application in this work is mainly
mid-rapidity. There,
initial state effects cannot explain high $p_T$ suppression, as
otherwise such
suppression should  also be observed in d-Au collisions, which is not
the case. High $p_T$ suppression has to arise from a
final state effect: jet energy loss\cite{Blaizot:2004px}. 
On top of experimental evidence for the suppression cited above, it has also
been proposed~\cite{wang3,gyu_vitev_wang} that azimuthal anisotropy at high
$p_T$ could also be explained by jet energy loss.  Induced gluon
bremsstrahlung, rather than elastic parton scattering, has been
identified to be the dominant
mechanism for jet energy loss~\cite{Gyu_plum,wgp}.  In the thermal medium, a
coherence effect, the Landau-Pomeranchuk-Migdal (LPM)~\cite{migdal}
effect, controls the strength of the bremsstrahlung emission.  Several
models of
jet quenching through gluon bremsstrahlung have been elaborated:
Baier-Dokshitzer-Mueller-Peign\'e-Schiff (BDMPS)~\cite{BDMPS,bdmps2001},
Gyulassy-Levai-Vitev (GLV)~\cite{GLV}, Kovner-Wiedemann (KW)~\cite{KW}
and Zakharov~\cite{zakharov1}.  There have also been phenomenological studies 
where different 
energy loss mechanisms were added onto Monte Carlo jet models  
\cite{Wang:1991xy,Wang:1996yh,jeon}
(see also \cite{Gyulassy:2003mc} for a recent review).
In
this paper, we will use the formalism developed by Arnold, Moore, and Yaffe
(AMY)~\cite{AMY}, which correctly treats the LPM effect (up to $O(g_s)$
corrections), to get the parton energy loss.  That model will be convolved
with a $(1+1)$D expanding QGP to obtain quantitative results on jet quenching.
Then, the electromagnetic signature of jet-plasma interactions will be
investigated.

This paper is organized as follows. In Sec.~II we give a qualitative
description of gluon radiation and briefly review the LPM effect
and the conditions for the LPM regime.  In Sec.~III, the AMY formalism is
presented and the main distinctions between the different treatments are
highlighted.  In the following sections, the complete space-time evolution of
the jet is calculated within a hydrodynamical model, and high $p_T$ spectra 
are obtained for pions (Sec.~IV) and real photons (Sec.~V), for
both RHIC and Large Hadron Collider (LHC) energies. Finally, Sec.~VI contains a
summary and conclusions.

\section{Qualitative Arguments}
\label{sec:qual}

 The high $p_T$
 data from various RHIC experiments (see Ref.~\cite{phenix_white} for a
 recent review) can be characterized by the following quantity:
 \be
 R_{AA} \equiv
 {(d N_{AA}/dy d^2p_T) \over \ave{N_{\rm coll}} (d N_{pp} /dy d^2p_T)}
 \approx C  \ \ \ \hbox{for}\ \ p_T \gsim 2\,\GeV \, ,
 \ee
 which represents the ratio of the number of events per unit rapidity
 and transverse momentum, scaled to the number expected based on
 proton-proton rates and the number of nuclear collisions occurring in
 the heavy ion collision.
 As indicated, this ratio is found experimentally to be approximately
 constant, with the constant $C$ ranging from about 0.2 to 0.6 depending
 on the centrality class and the rapidity $y$.
 The $pp$ spectrum at mid-rapidity
 can be conveniently parametrized by \cite{Adler:2003au}
 \be
 {d N_{pp} \over dy d^2p_T}
 =
 A \left( p_0\over p_0 + p_T \right)^n \, ,
\label{eq:init_spect}
 \ee
 where $A$ is a normalization constant, $p_0 = 1.2\,\GeV$ and $n = 10$.
\begin{figure}[t]
\epsfxsize=0.6\tw
\epsfysize=0.4\tw
\centerline{\epsfbox{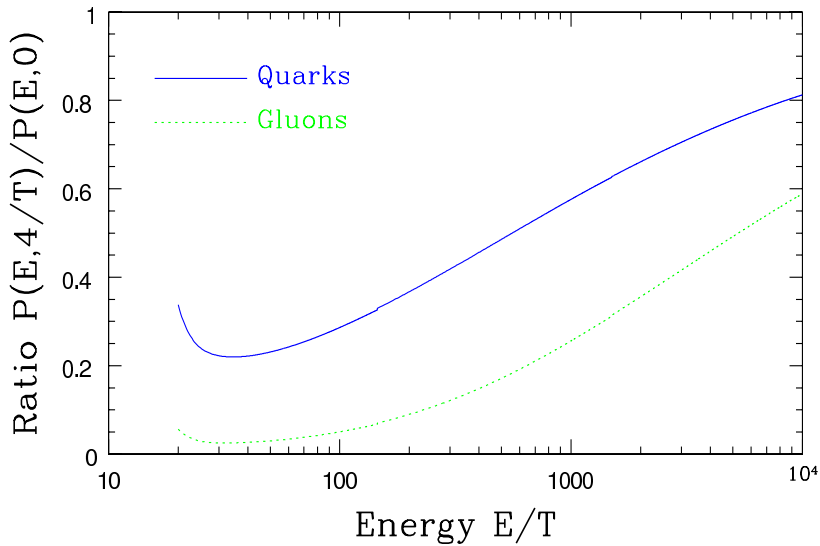}}
\caption{(Color online)
The ratio of the final and the initial parton spectrum calculated in
Ref.~\cite{Jeon-Moore} using the leading order thermal QCD result.
}
\label{fig:fig4}
\end{figure}
 In a previous paper \cite{Jeon-Moore} two of us showed that the 
 leading order thermal QCD
 result, shown in Fig.~\ref{fig:fig4}, can qualitatively explain the fact 
 that $C$ is roughly constant over a finite kinematical window. 
 In this section, we use the simple gluon radiation spectrum used
 by Baier {\it et.\ al.\ } \cite{BDMPS} to illustrate the main physical
 features of this calculation.

 As previously stated, the dominant energy loss mechanism in medium is
 the radiation of gluons.  We will express the rate of radiation of
 gluons of energy $\omega$, per unit time and $\omega$ interval,
 as $\Gamma\equiv dI/d\omega dt$.
 To understand the relationship between the behavior of $R_{AA}$ and
 $\Gamma$,  we note that, given $\Gamma$, the parton
 distribution function $P(p,t)$ (which gives the probability that there
 will be a parton of energy $p$ at time $t$) should satisfy a rate equation,
 \be
 {dP(p, t)\over dt}
 =
 \int_{-\infty}^\infty d\omega\, P(p+\omega, t)\, \Gamma(p+\omega, \omega, t)
 -
 P(p, t)\, \int_{-\infty}^\infty d\omega\, \Gamma(p, \omega, t) \, ,
 \ee
 provided that conversion between different species of partons is
 unimportant.
 If the rate $\Gamma(p,\omega,t)$ has a weak dependence on $p$
 (an assumption we make here for illustration but will not use in our
 work in subsequent sections),
 then an approximate solution can be written as
\be
P(p, t) \approx
\int d\epsilon\, D(\epsilon, p, t)\, P_0(p + \epsilon) \, ,
\label{eq:simple_sol}
\ee
where
\be
D(\epsilon, p, t)
& = &
\exp\left[-\int_{-\infty}^{\infty} d\omega\, J(p, \omega, t)\right]
\sum_{n=0}^\infty
{1\over n!}
\left[
\prod_{i=1}^n\int_{-\infty}^\infty d\omega_i\,
J(p+\omega_i,\omega_i, t)
\delta\left(\epsilon - \sum_{i=1}^n\omega_i\right)
\right]
\label{eq:Depsilon}
\ee
with
\be
J(p, \omega, t) \equiv \int_0^t dt'\, \Gamma(p, \omega, t') \,.
\ee
A similar expression is written down in Ref.~\cite{bdmps2001}.
%The differences between the current expression and the one in~\cite{bdmps2001}
%are that in  Eq.~(\ref{eq:Depsilon}), the argument of
%function $J$ under the summation sign
%has $p+\omega_i$ whereas in Ref.\cite{bdmps2001} this was just $p$ and
% Guy had to cut this because we realized that the approximation
% needed in the derivation is that $J$ is essentially independent of
% this change.
The main difference between the current expression and the one
in~\cite{bdmps2001} is that
the range of integration in the above equations is from $-\infty$ to
$\infty$: The negative $\omega$ corresponds to the absorption of thermal
partons and the range $\omega > p$ corresponds to annihilation of a parton
with a thermal anti-parton.

Following \cite{bdmps2001}, we let
\be
{P_0(p+\epsilon)\over P_0(p)}
\approx \exp(-n\epsilon/p) \,,
\label{eq:exp_approx}
\ee
which allows us to write the nuclear modification factor as
\be
R_{AA}(p)
& = &
\int d\epsilon \, D(p,\epsilon,t) {P_0(p+\epsilon)\over P_0(p)}
\non
& \approx &
\int d\epsilon \, D(p,\epsilon,t) \, \exp(-{n\epsilon/p})
%\non
%& = &
%\exp\left[-\int d\omega\, J(p, \omega, t)\right]
%\sum_{n=0}^\infty
%{1\over n!}
%\left[
%\prod_{i=1}^n\int d\omega_i\,
%J(p+\omega_i,\omega_i, t)
%e^{-{n\over p}\omega_i}
%\right]
\non
& = &
\exp
\Bigg[
-\int_{-\infty}^\infty d\omega\, J(p,\omega, t)
+ \int_{-\infty}^\infty d\omega\, J(p+\omega, \omega, t)e^{-\omega/(p/n)}
\Bigg] \, .
\ee
For this to be approximately constant, or at least have a weak dependence on
$p$, the exponent
\be
A(p,t,n) =
-\int_{-\infty}^{\infty} d\omega\, J(p, \omega, t)
+ \int_{-\infty}^\infty d\omega\, J(p+\omega, \omega, t)e^{-\omega/(p/n)}
\label{eq:exponent}
\ee
must have weak dependence on $p$,
at most equaling $\beta \ln p$ with $\beta < 1$.

Even without knowing the explicit form of the $J$ function, some insight can be
gained 
%there are a few
%things we can gain 
from this expression.  First, for the second
integral to be well defined in the $\omega < 0$ range,
the Boltzmann factor corresponding to the thermal partons must compensate
the $e^{-\omega/(p/n)}$ factor.
Secondly, because of the $e^{-\omega/(p/n)}$ factor, in general the loss term
dominates and hence $R_{AA}(p) < 1$.
Third, as $p\to \infty$, the compensation between the two terms in $A$ becomes
complete and $R_{AA} \to 1$.

To be more specific, the precise form of the $J$ function needs to be known.
The energy loss of an energetic parton in a dense medium mainly proceeds via
 bremsstrahlung of soft gluons.  The form of the energy loss (the $J$
 function above)
 depends crucially on whether bremsstrahlung off many scatterers are
 coherent or incoherent. This is determined by the relative size of
 the mean free path $\lambda = 1/n\sigma$ and
 the coherence length \cite{Baier:2000mf},
 \be
 l_{\rm coh} = \sqrt{\lambda\omega\over \mu^2}
 \label{eq:l_coh}
 \ee
 where $\mu$ is the typical size of the soft momentum exchange
 and $\omega$ is the energy of the emitted gluon.
 When $l_{\rm coh} \ll \lambda$, then one is in the extreme
 Bethe-Heitler regime where
 the energy loss per unit length is proportional to the incoming energy.
 On the other hand, when $l_{\rm coh} \gg \lambda$, then one is in the extreme
 Landau-Pomeranchuk-Migdal (LPM) regime where the energy loss per unit
 length is proportional to the square-root of the incoming energy.

 The typical mean free path for soft scattering in a hot medium can be
 estimated as follows.
 The density of scatterers according to the thermal distribution is
 $n \sim T^3$.  The typical soft ($p_{\rm exch} \sim gT$)
 scattering cross-section is given by
 $\sigma \sim g^2/T^2$.  Hence
 \be
 \lambda \sim 1/g^2 T \, .
 \ee
 According to the above estimate of the coherence length, the LPM effect
 becomes relevant when $l_{\rm coh} \gsim \lambda$ or
 \be
 {\omega\over \mu^2} \gsim \lambda \, ,
 \ee
 with $\mu \sim gT$ and $\lambda \sim 1/g^2T$.  This yields the condition
 \be
 \omega \gsim T \, .
 \ee
 Defining $E_{\rm LPM} = \lambda \mu^2$, one can then conclude that
 for the emitted gluon energy in the range of
 $0 < \omega < E_{\rm LPM}$, energy loss is governed by the
 Bethe-Heitler limit
 \be
 \omega {dI\over d\omega dz} \simeq {\alpha_s\over \pi}N_c {1\over
 \lambda} \, ,
 \ee
 and for $\omega > E_{\rm LPM}$, the energy loss is governed by the LPM
 limit \cite{Baier:2000mf},
 \be
 \omega {dI\over d\omega dz} \simeq {\alpha_s\over \pi}N_c
 \sqrt{\mu^2\over \lambda\omega} \, .
 \ee

 For $\omega > E_{\rm LPM}(L/\lambda)^2$,
 the coherence length exceeds the length of
 the medium.  In this case, effectively only a single scattering can occur
 and the radiation spectrum goes back to
 \be
 \omega {dI\over d\omega dz} \simeq {\alpha_s\over \pi}N_c {1\over L}\, .
 \ee
 In this paper, we will only consider the case
 \be
 0 < E_{\rm LPM} \leq E_0 < E_{\rm fact}
 \ee
 assuming that $L/\lambda \gg 1$.
 Here $E_0$ denotes the original energy of the parton
 and we defined $E_{\rm fact} = E_{\rm LPM}(L/\lambda)^2$.

In order to use Eq.~(\ref{eq:exponent}), in addition to the above forms of
gluon spectrum for $\omega > 0$,
we also need to know the form of $J(p,\omega,t)$ for $\omega < 0$.
For simplicity, we take the spectrum for $\omega <0$ to be the same as the
Bethe-Heitler spectrum and multiply it by the Boltzmann factor to take into
account that absorption of high energy gluons from the thermal medium
is Boltzmann suppressed:
\be
{dI\over d\omega dz}
= {\alpha_s N_c\over \pi\lambda} {1\over |\omega|}\, e^{-|\omega|/T}
\label{eq:I_of_negative_w} \, .
\ee

\begin{figure}[t]
\epsfxsize=0.6\tw
\epsfysize=0.4\tw
\centerline{\epsfbox{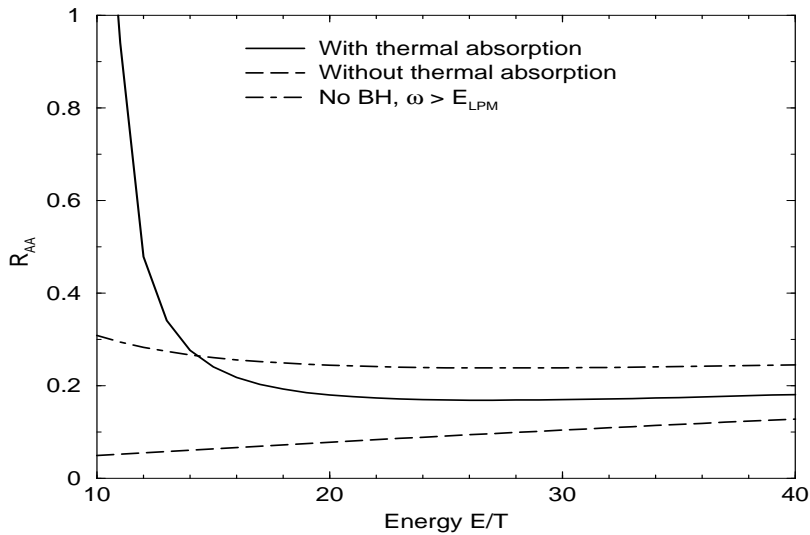}}
\caption{
Approximate $R_{AA}(p)$
calculated with Eq.~(\protect\ref{eq:exponent}) following the approach of
BDMPS up
to the maximum momentum of $E_{\rm fact} = 16\,\GeV$.
The solid line includes the contribution from absorption of thermal gluons
whereas the broken line only includes the emission.
The dashed-dotted line is calculated with only the LPM part of the gluon
spectrum.
The divergence at $p = 10T$ is an artifact of the approximation.
}
\label{fig:bdmps_raa_part}
\end{figure}

\begin{figure}[t]
\epsfxsize=0.6\tw
\epsfysize=0.4\tw
\centerline{\epsfbox{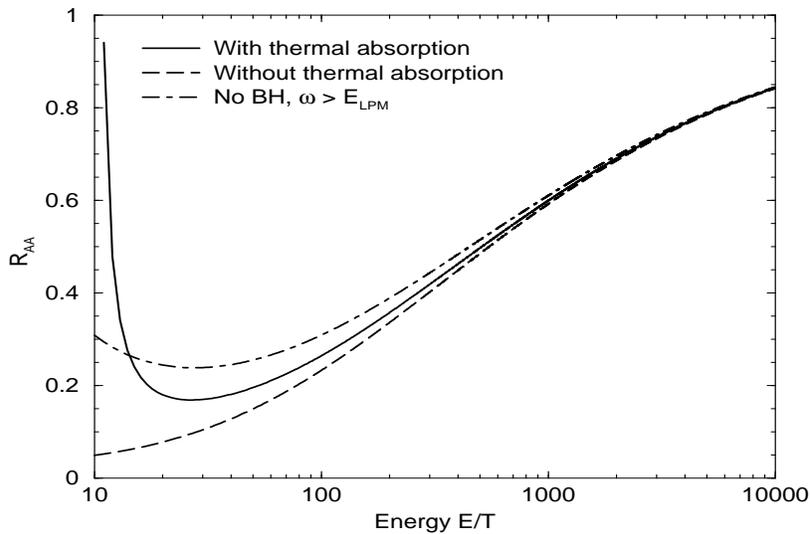}}
\caption{
The same as Fig.\protect\ref{fig:bdmps_raa_part} but showing a larger energy
range.
}
\label{fig:bdmps_raa}
\end{figure}

For illustration, we set
$E_{\rm LPM} = 1\,\GeV$, $\lambda = 1\,\fm$
and $T=400\,\MeV$ and evolve the system until $L = ct = 4\,\fm$.
We have also set the upper limit of the momentum integral to $p$.
The above simple gluon spectrum then yields the $R_{AA}(p)$
curves shown in Fig.\ref{fig:bdmps_raa_part}.
The solid curves in the figures are calculated including the absorption
contribution and the broken curves are calculated without the absorption
part. Also shown are the dot-dashed curves which are calculated with only
the $\omega > E_{\rm LPM}$ part of the gluon spectrum omitting the
Bethe-Heitler part.
In Fig.\ref{fig:bdmps_raa},
we show a momentum range far exceeding the factorization
energy to compare with the calculation in Ref.~\cite{Jeon-Moore}.

In Ref~\cite{Jeon-Moore}, two of us
have calculated the same ratio of the final and the
initial spectrum using the resummation method developed by AMY.
The shape and the trend of the curve calculated with the thermal absorption
in Figs.\ref{fig:bdmps_raa_part} and \ref{fig:bdmps_raa} closely match up
with what we have previously calculated in Ref~\cite{Jeon-Moore}.
The experimentally obtained $R_{AA}$ from STAR and PHENIX from RHIC
experiments also roughly has the same shape and 
trend~\cite{Adams:2003im,phenix2}.

There are two major
points we would like to make with this simple calculation.
One is
the importance of the {\em absorption} of the thermal gluons.
This is also closely related to the appearance of a momentum scale $p = nT$.
Near this scale,
absorption of thermal gluons is very efficient in changing the
shape of the final spectrum.
Comparing the two curves in Fig.\ref{fig:bdmps_raa_part} and also in
Fig.\ref{fig:bdmps_raa}, it is clear that the roughly flat shape within the
limited momentum range shown in Fig.\ref{fig:bdmps_raa_part} is due to the
absorption of thermal gluons.  Furthermore, the efficient absorption near
$p=nT$ causes $R_{AA}$ to have a larger value near this point.
Without the absorption, $R_{AA}$ is a monotonically increasing function of
$p$.  In the experimentally measured $R_{AA}$, one does observe such a
change in slope near $p_T = 3\,\GeV$. This is usually attributed to the
Cronin effect.  However, it could very well be due to the absorption of
thermal gluons.

The importance of gluon absorption will arise whenever the slope of the
spectrum, $d\ln P/dp$, is comparable with the thermal value $1/T$.  When
these are equal, detailed balance ensures that the more numerous lower
energy particles keep re-populating the higher energy ones.  The reason
the spectrum in Fig.~\ref{fig:bdmps_raa_part} rises to 1 at $10T$ is
because we took
the initial particle distribution to be a strict power law, $\sim
p^{-n}$, which becomes steeper than the thermal spectrum below $p=nT$.
More realistic initial distributions, as in
Eq.~(\ref{eq:init_spect}), will not display this behavior; nevertheless,
absorption is important in the lower energy region.

Another point to make is the importance of the Bethe-Heitler part of the
radiation spectrum.  The effect of neglecting the Bethe-Heitler part of the
spectrum becomes justifiable only around $p \sim 100T$. Therefore,
calculations with only the LPM part of the spectrum appear unrealistic for
RHIC energies.


\section{The formalism}
\label{lab:formalism}

\begin{figure}[t]
\epsfxsize=0.6\tw
\centerline{\epsfbox{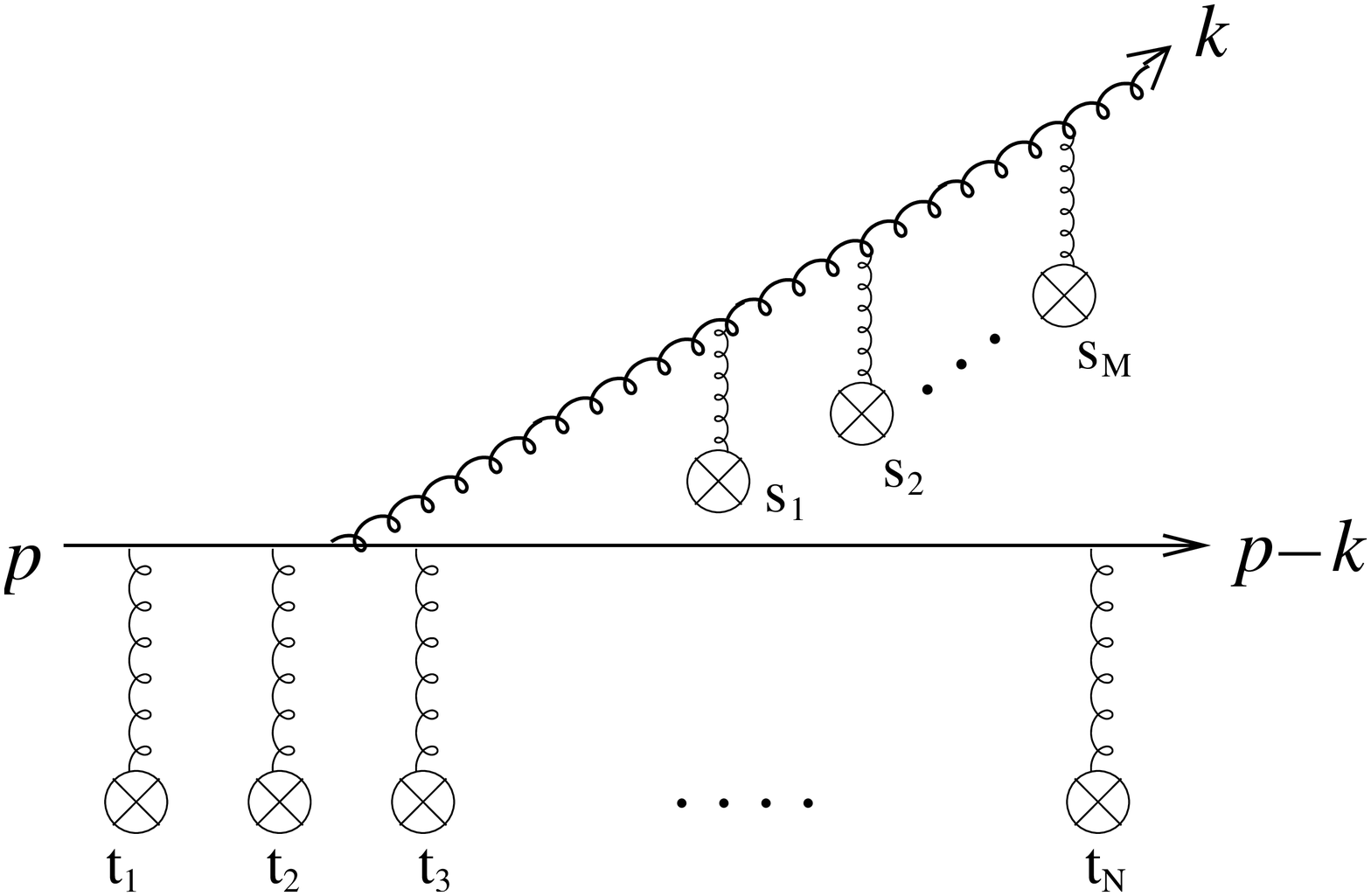}}
\caption{A typical bremsstrahlung diagram that needs to be resummed.}
\label{fig:the_big_picture}
\end{figure}

 In a seminal series of papers,
 Baier, Dokshitzer, Mueller, Peign\'e and Schiff identified the diagrams that
 contribute to the leading order energy loss in a hot QGP medium.
 A typical diagram is shown in Fig.\ref{fig:the_big_picture}.
To compute the energy loss rate, one must deal with the following issues:
\begin{enumerate}
\item
How are these diagrams to be resummed?
\item
How is the resulting equation to be solved?
\item
How does one describe the {\em time evolution} of the energy loss?
\end{enumerate}
 We now summarize the approach taken by Arnold, Moore
 and Yaffe (AMY).  The ingredients discussed in this section can be found in
 Ref.~\cite {Jeon-Moore}. We shall also briefly compare with other approaches.

\begin{figure}[t]
\epsfxsize=0.6\tw
\centerline{\epsfbox{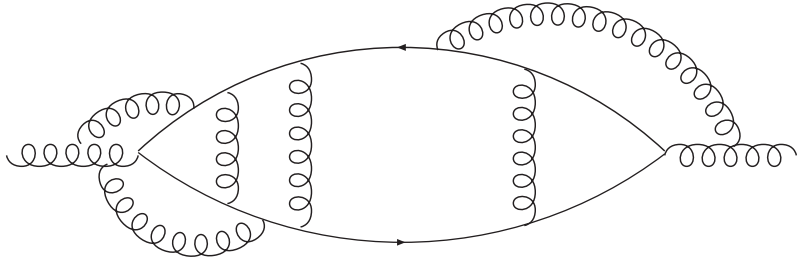}}
\caption{A typical gluon self-energy diagram that needs to be resummed.}
\label{fig:g_self_energy}
\end{figure}

 In the diagram shown in Fig.~\ref{fig:the_big_picture}, a single
 wiggly line connected to a single
 $\otimes$ is not a gluon propagator. Rather, it just denotes
 that a quark has interacted with the soft background field at a certain
 time $t_i$ or $s_i$.
 To get the gluon energy spectrum at the final time, we need to sum all the
 diagrams of the type shown in Fig.~\ref{fig:the_big_picture}, then square and
 average over the soft field configurations.
 This procedure leads to the conclusion that the scattering rate can be
 obtained by calculating the imaginary part of the gluon self-energy in the
 soft thermal background.  A typical gluon self-energy in such a situation
 is shown in Fig.~\ref{fig:g_self_energy}.

The result of resummation is summarized in the following formulas.
The transition rates for various emission processes are given by
\begin{eqnarray}
\label{eq:dGamma}
\frac{d\Gamma(p,k)}{dk dt} & = & \frac{C_s \gs^2}{16\pi p^7}
        \frac{1}{1 \pm e^{-k/T}} \frac{1}{1 \pm e^{-(p-k)/T}}
\times
%\nonumber \\ && \times
\left\{ \begin{array}{cc}
        \frac{1+(1{-}x)^2}{x^3(1{-}x)^2} & q \rightarrow qg \\
        N_{\rm f} \frac{x^2+(1{-}x)^2}{x^2(1{-}x)^2} & g \rightarrow qq \\
        \frac{1+x^4+(1{-}x)^4}{x^3(1{-}x)^3} & g \rightarrow gg \\
        \end{array} \right\} \times \nonumber \\ && \times
\int \frac{d^2 \h}{(2\pi)^2} 2 \h \cdot {\rm Re}\: \F(\h,p,k) \, .
\end{eqnarray}
Here $C_s$ is the quadratic Casimir relevant for the process (in QCD,
$4/3$ for processes involving a quark and 3 for the pure glue process),
and $x\equiv k/p$ is the momentum fraction in the gluon (or the
quark, for the case $g \rightarrow q\bar{q}$).
The factors $1/(1\pm e^{-k/T})$ are Bose stimulation or Pauli blocking
factors for the final states, with $-$ for bosons and $+$ for fermions.
$\h \equiv \p \times
\bfk$ determines how non-collinear the final state is; we treat it as
parametrically $O(g T^2)$ and therefore small compared to $\p\cdot \bfk$.
Therefore it can be taken as a two-dimensional vector in the transverse
space.
$\F(\h,p,k)$ is the solution of the following integral equation:
\begin{eqnarray}
2\h = &&\hspace{-0.14in}
        i \delta E(\h,p,k) \F(\h) + g_s^2 \int \frac{d^2 \q_\perp}{(2\pi)^2}
C(\q_\perp)
\times
%\nonumber \\ && \hspace{1.5in}\times
	\Big\{ (C_s-C_{\rm A}/2)[\F(\h)-\F(\h{-}k\,\q_\perp)]
        \nonumber \\ && \hspace{2.65in}
        + (C_{\rm A}/2)[\F(\h)-\F(\h{+}p\,\q_\perp)]
        \nonumber \\ && \hspace{2.65in}
        +(C_{\rm A}/2)[\F(\h)-\F(\h{-}(p{-}k)\,\q_\perp)] \Big\} \,.
\label{eq:integral_eq1}
\end{eqnarray}
Here $C(\q_\perp)$ is the differential rate to exchange transverse (to
the parton) momentum $\q_\perp$.  In a hot thermal medium, its value at
leading order in $\alpha_{\rm s}$ is \cite{AGZ}
\begin{equation}
\label{eq:Cq}
C(\q_\perp) = \frac{\mD^2}{\q_\perp^2(\q_\perp^2{+}\mD^2)} \, ,
\quad
\mD^2 = \frac{\gs^2 T^2}{6} (2 N_{\rm c} {+} N_{\rm f}) \, .
\end{equation}
The energy difference between the final and the initial states is given by
\be
\delta E(\h,p,k) &&\hspace{-0.14in}
        = \frac{\h^2}{2pk(p{-}k)} + \frac{m_k^2}{2k} +
        \frac{m_{p{-}k}^2}{2(p{-}k)} - \frac{m_p^2}{2p} \, ,
\label{eq:integral_eq2}
\ee
where $m^2$ are the medium induced thermal masses, equal to $m_D^2/2$ for
a gluon and $C_f \gs^2 T^2/4 = \gs^2 T^2/3$ for a quark.  For the case
of $g\rightarrow qq$, $(C_s-C_{\rm A}/2)$ should appear as the
prefactor on the term containing
$\F(\h-p\,\q_\perp)$ rather than $\F(\h-k \,\q_\perp)$.

Next, we use these expressions to evolve the hard gluon distribution
$\Pg (p,t=0)$ and the hard quark plus antiquark distribution
$\Pq (p,t=0)$ with time, as they traverse the medium.
The joint equations for $\Pq$ and $\Pg$ are
\begin{eqnarray}
\frac{d\Pq (p)}{dt} & = & \int_k \!
        \Pq (p{+}k) \frac{d\Gamma^q_{\!qg}(p{+}k,k)}{dkdt}
        -\Pq (p)\frac{d\Gamma^q_{\!qg}(p,k)}{dkdt}
%        \nonumber \\ && \quad
        +2 \Pg (p{+}k)\frac{d\Gamma^g_{\!q \bar q}(p{+}k,k)}{dkdt}
        \, , \nonumber \\
\frac{d\Pg (p)}{dt} &  = & \int_k \!
        \Pq (p{+}k) \frac{d\Gamma^q_{\!qg}(p{+}k,p)}{dkdt}
        {+}\Pg (p{+}k)\frac{d\Gamma^g_{\!\!gg}(p{+}k,k)}{dkdt}
        \nonumber \\ && \;
        -\Pg (p) \left(\frac{d\Gamma^g_{\!q \bar q}(p,k)}{dkdt}
        + \frac{d\Gamma^g_{\!\!gg}(p,k)}{dkdt} \Theta(2k{-}p) \!\!\right) ,
\label{eq:Fokker}
\end{eqnarray}
where the $k$ integrals run from $-\infty$ to
$\infty$.  The integration range with $k<0$ represents absorption of
thermal gluons from the QGP; the range with $k>p$ represents
annihilation against an antiquark from the QGP, of energy $(k{-}p)$.
In writing \eqn{eq:Fokker}, we used
$d\Gamma^g_{\!gg}(p,k)=d\Gamma^g_{\!gg}(p,p{-}k)$ and similarly for
$g\rightarrow qq$; the $\Theta$ function in the loss term for $g
\rightarrow gg$ prevents double counting of final states.  Since
bremsstrahlung energy loss involves only small $O(\gs T/p)$ changes to the
directions of particles, these equations can be used for the momentum
distributions in any particular direction.  For a single initial hard
particle, they can be viewed as Fokker-Planck equations for
the evolution of the probability distribution of the particle energy and
of accompanying gluons.  These expressions depend at several points on
$g_{\rm s}^2$ or $\alpha_{\rm s}$.  When evaluating them numerically, we
have used $\alpha_{\rm s}=0.3$.

We now briefly compare our approach with others used in the
literature.
As this paper is not intended to be a review, this cannot and will not be
a complete comparison with all other jet energy loss calculations.

As was mentioned at the beginning of this section, most jet quenching
calculations start with the problem of resumming all diagrams of the form
shown in Fig.\ref{fig:the_big_picture}.
In a series of papers, U.~Wiedemann {\it et al.\ } have shown that
BDMPS, Zakharov,
GLV and also the eikonal approach taken by Wiedemann and Kovner
are all related to each other.
In these approaches, the problem of resumming the diagrams is solved in
position space assuming {\em static scatterers}, although the methods of
solving the problem differ.
Since a thermal medium consists of dynamic, moving scatterers,
temperature was introduced only as the controlling variable for the mean
free path in these approaches.
The evolution of the initial parton distribution is then achieved through
Eq.~(\ref{eq:simple_sol}) with the important restrictions that only the {\em
emission} of gluons was considered with (Wiedemann) or without (BDMPS)
restriction on the kinematic upper limit of the emitted energy.
There is also an approximation that ignores the emitted energy in the
gain rate ($\Gamma(p+\omega,\omega,t)\to \Gamma(p,\omega,t)$ in
Eq.~(\ref{eq:simple_sol})).

Our approach differs with the above ones in three major ways.
The biggest difference is that our calculation is completely thermal and
hence the scatterers are all dynamic.  In our calculation,
temperature enters through the thermal phase space of the initial and the
final particles and there is no assumption of the form of the
elementary cross-section.  All are calculated completely within the
framework of hard thermal loop resummed leading order thermal QCD.
Hence, gain or loss due to the absorption of thermal partons $(\omega < 0)$
as well as the loss process of pair annihilations with the thermal partons
$(\omega > p)$ are fully included in our calculation. These are missing in
the approaches mentioned above.  In Ref.~\cite{Wang-Wang}, thermal absorption
and stimulated emission were introduced in the framework of GLV.
This was done only up to the first order in the opacity expansion
without including the annihilation process with thermal partons.
Second, instead of approximately solving for the transition rates and then
using Eq.~(\ref{eq:simple_sol}), we explicitly solve for both the transition
rates and the coupled rate equations of hard quarks and gluons.  There
are also no approximations about whether the Bethe-Heitler or LPM regime
is relevant; the transition between these extremes is handled correctly.

One limitation of the current approach as compared to the previous ones is
that the transition rates are calculated in momentum space assuming
the thermodynamic limit.  That is, our approach assumes that the
high energy parton experiences a uniform medium on the time scale of
the formation time of the emitted radiation.  Hence, our approach is
limited to momenta less than the factorization energy.
With $\mu \approx 0.5\,\GeV$, $L \approx 5\,\fm$ and $\lambda
\approx 1\,\GeV$, this limits the momenta to the region $p < 30\,\GeV$.
This is not a big problem at RHIC energies, where $\sqrt{s}$ is only
$200\,\GeV$, and it also covers the $p_T$ acceptance of the ALICE
detector at the LHC.

%%%%%%%%%%%%%%%%%%%%%%%%%%%%%%%%%%%%%%%%%%%%%%%%%%%%%%%%%%%%%%%%%%%%%%%%%%%%
\section{Pion Production}
\label{pion}
%%%%%%%%%%%%%%%%%%%%%%%%%%%%%%%%%%%%%%%%%%%%%%%%%%%%%%%%%%%%%%%%%%%%%%%%%%%%

The goal of this section is to use the formalism explained in
the previous section to calculate the neutral pion spectrum in heavy
ion collisions.  Our approach to this problem relies on the fact that for
hard spectra, the $AA$ collision can be regarded as a collection of binary
collisions.  In this way of formulating the problem, the $AA$ spectrum is given
by the convolution of the elementary $pp$ spectrum with geometrical and
in-medium factors.

Up to suppressed corrections, the $\pi^0$ cross-section in an N-N
collision factorizes and is given by~\cite{owens}
\begin{equation}
\label{pi0_nn}
E_{\pi}\frac{d\sigma_{NN}}{d^3p_{\pi}}=\sum_{a,b,c,d}
\int dx_a dx_b g(x_a,Q)g(x_b,Q)
K_{jet}\frac{d\sigma^{a+b\rightarrow c+d}}{dt}\frac{1}{\pi z} D_{\pi^0/c}(z,
Q^{'})
\end{equation}
where $g(x,Q)$ is the parton distribution function (PDF) in a nucleon,
$D_{\pi^0/c}(z, Q^{'})$ is the pion fragmentation function,
$\frac{d\sigma^{a+b\rightarrow c+d}}{dt}$ is the parton-parton cross-section at
leading order and the factor $K_{jet}$ accounts for higher
order effects (where ``jet'' here means a fast parton having $p^{jet}_T
\gg$ 1 GeV).  According to~\cite{barnafoldi}, $K_{jet}$ is almost
$p^{jet}_T$ independent at RHIC energies.  We will use
$K_{jet}\sim$ 1.7 for RHIC and 1.6 for the LHC, based on their results.

For all our calculations,
we set the factorization scale ($Q$) and the fragmentation
scale ($Q^{'}$) equal to $p_T$.
We take the CTEQ5 parton distribution function
{}~\cite{cteq5} and the $\pi$ fragmentation function extracted from $e^+e^-$
collisions \cite{binnewies}.

\begin{figure}[t]
\centerline{
\psfig{file=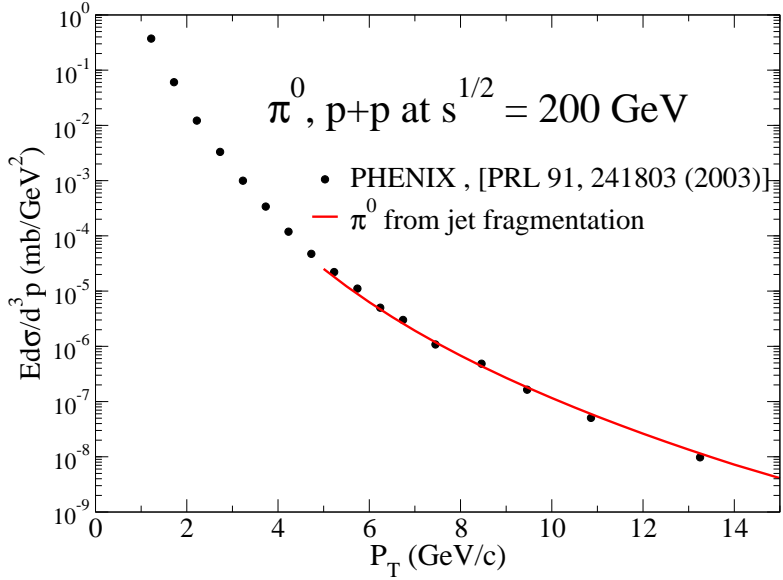,width=8.5cm,angle=-90}
}
\caption{(Color online) Neutral pion spectra in $pp$ collisions at RHIC.
The data points are
from PHENIX and the solid line is the calculated result from jet
fragmentation.}
\label{pi0}
\end{figure}

Fig.~\ref{pi0} shows our calculation for the spectrum of high $p_T$
neutral pions in $pp$ collisions at RHIC,
compared to the PHENIX result~\cite{phenix}.  One can readily see that
our calculation reproduces the data in
the region where jet fragmentation is expected to be the dominant mechanism of
particle production ($p_T >$ 5 GeV/c)~\cite{fries}.

 To obtain the high $p_T$ $\pi^0$ cross-section in $AA$ collisions, we
 must modify the $pp$ calculation in two ways.  First, the PDF of a
 nucleus differs from that of a proton;
 \begin{equation}
 g_A(x_a,Q)=g(x_a,Q) R_A(x_a,Q) \, ,
 \end{equation}
 where the nuclear modification of the structure function $R_A$ takes into
 account shadowing and anti-shadowing.
 In this paper, we use $R_A(x_a, Q)$ as parametrized by Eskola
 {\it et al.}~\cite{eskola}.

 Second, we must account for the energy loss of the parton between its
 production in the hard initial scattering event and its hadronization.
 We will assume that a jet fragments only outside the medium, as
 may be justified by estimating the formation time of a pion
 with the typical observed energy.
 In Ref.\cite{wang}, the formation time of an emitted pion with energy
 $E_\pi$ is estimated to be
 $\sim R_{\pi}\frac{E_{\pi}}{m_{\pi}}$, where
 $R_{\pi}\sim 1\,\fm$ is the size of the pion.
 For a 10 GeV pion, this gives $\tau_f\sim$ 35-70 fm/c. This is much
 longer than the path of a jet in the hot matter created by two colliding
 nuclei.  Therefore, one should consider production, energy loss in
 medium, and fragmentation to occur sequentially.

 We will assume that the fragmentation at the edge of the QGP involves
 the usual vacuum fragmentation function.  The medium effect is then to
 reduce the 
 parton energy by an amount determined by the Fokker-Planck equation
 previously presented, Eq.~(\ref{eq:Fokker}).
 This is most conveniently written by defining a new medium-inclusive
 effective fragmentation function,
\begin{equation}
\tilde D_{\pi^0,c}(z,Q;\bfr,\bfn) = \int dp_f\frac{z'}{z} \left(
	P_{\!q\bar{q}/c}(p_f;p_i) D_{\pi^0/q}(z',Q) +
	P_{\!g/c}(p_f;p_i) D_{\pi^0/g}(z',Q) \right) \, ,
\label{eq:Dtilde}
\end{equation}
 where $z=p_T/p_i$ and $z'=p_T/p_f$.  $P_{\!q\bar{q}/c}(p_f;p_i)$ and
 $P_{\!g/c}(p_f;p_i)$ represent the solution to Eq.~(\ref{eq:Fokker}),
 which is the probability to get a given parton with final momentum $p_f$
 when the initial condition is a particle of type $c$ and momentum
 $p_i$.

 The quantities $P_{\!q\bar{q}/c}$ and $P_{\!g/c}$ in Eq.~(\ref{eq:Dtilde})
 depend implicitly on the path
 length the initial parton must travel through the medium and the
 temperature profile along that path.  This is not the same for all
 jets, because it depends on the location where the jet is produced and
 on the direction the jet propagates.  Therefore, one must convolve this
 expression over all transverse positions $\bfr_\perp$ and directions
 $\bfn$.  Since the number of jets at $\bfr_\perp$
 is proportional to the number of binary collisions, the probability
 is proportional to the the product of the thickness functions
 of the colliding nuclei at $\bfr_\perp$.
 For central collisions where the impact parameter $b\approx 0$, we get
\be
\calP(\bfr_\perp)
\propto
T_A(\bfr_\perp)\, T_B(\bfr_\perp) \, .
\ee
For a hard sphere which we use for simplicity,
this probability is
\be
\calP(\bfr_\perp) =
{2\over \pi R_\perp^2}
\left(1 - {r_\perp^2\over R_\perp^2}\right)
\, \theta(R_\perp - r_\perp) \, ,
\label{Pr_dep}
\ee
which is normalized to yield
$
\int d^2 r_\perp\, \calP(\bfr_\perp) = 1
$.  Since the direction of the jet is fixed by the pion direction
($\bfn=\frac{{\bf p_{\pi}}}{|{\bf p_{\pi}}|}$),
the final in-medium modified fragmentation function is
\be
\label{D_conv}
\tilde D_{\pi^0/c}(z,Q)
=
\int d^2 r_\perp\,
\calP(\bfr_\perp)
\tilde{D}_{\pi^0/c}(z, Q, \bfr_\perp, \bfn)\, .
\ee
The $AA$ spectrum is now given by
\be
\label{eq:pi0_AA}
\frac{dN_{AA}}{dyd^2p_{T}}=\frac{\ave{N_{\rm coll}}}
{\sigma_{in}}\sum_{a,b,c,d}
\int dx_a dx_b\,
g_A(x_a,Q)g_A(x_b,Q)
K_{jet}\frac{d\sigma^{a+b\rightarrow c+d}}{dt}
\frac{\tilde D_{\pi^0/c}(z, Q)}{\pi z} \, ,
\ee
where $\ave{N_{\rm coll}}$
is the average number of binary collisions
and $\sigma_{in}$ is the inelastic nucleon-nucleon cross-section.

\begin{figure}[t]
 \epsfxsize=0.6\tw
 \centerline{ \epsfbox{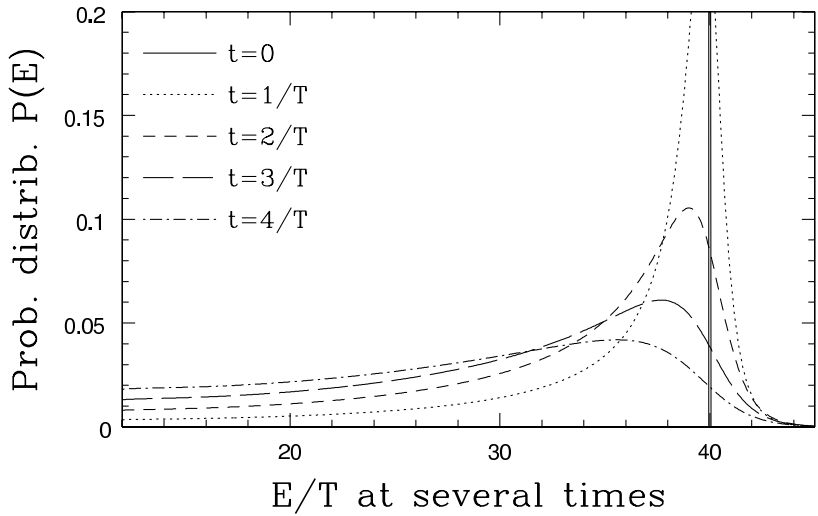} }
 \caption{The evolution of ${\cal P}(\Delta E|E_c)$ as a function of time
 (passage length).}
 \label{fig:evolv_40}
\end{figure}

Since Eq.~(\ref{eq:pi0_AA}) is expressed in terms of probability
distributions, it is straightforward to evaluate it using the Monte-Carlo
method.  One complication in doing so is that,
while the parton traverses the medium, the medium also evolves.
Therefore at each time step of solving Eq.~(\ref{eq:Fokker}),
the temperature must be adjusted to the local environment.  We assume 
that the medium expands only in the longitudinal direction, based on the 
following reasoning.  The low $p_T$ neutral pion spectrum at RHIC is
well reproduced by a hydrodynamical model incorporating transverse expansion, 
while the model fails for $p_T >$  3 GeV, suggesting that high-$p_T$ pions 
mainly come from jet fragmentation~\cite{niemi}. The transverse 
expansion will have two effects on jet energy loss.  First, the
expanding geometry will increase the
duration of parton propagation. However, the same expansion will make for a
falling parton density along the path. Those two effects partially compensate
each other and the energy loss is just about the same as in the case without
transverse expansion \cite{gvwh2002}.  In 1-D Bjorken
expansion~\cite{Bjorken}, the temperature evolves as
\be
\label{T1}
T=T_i\left(\frac{\tau_i}{\tau}\right)^{1/3} \, .
\ee

In the original Bjorken model, the transverse density is assumed to be
constant.  Since a nucleus does have a transverse density profile,
it is more realistic to assign the initial
temperatures in the transverse direction according
to the local density so that~\cite{prlphoton,prcdilep}
\be
\label{T2}
T(r,\tau_i)=T_i\left[2\left(1-\frac{r^2}{R_{\perp}^2}\right)\right]^{1/4}
\, .
\ee
Putting Eqs.(\ref{T1}) and (\ref{T2})
together, we get the temperature evolution of a QGP expanding in 1-D
as
\be
\label{Tevolution}
T(r,\tau)=T_i\left(\frac{\tau_i}{\tau}\right)^{1/3}
\left[2\left(1-\frac{r^2}{R_{\perp}^2}\right)\right]^{1/4} \, .
\ee

The jet evolves in the QGP medium until it reaches
the surface or until the temperature reaches the transition
temperature $T_c$.  In our calculation,
we assume a first-order phase transition and use
\be
f_{QGP}=\frac{1}{r_d-1}\left(\frac{r_d\tau_f}{\tau}-1 \right)
\label{fmix}
\ee
as the fraction of the QGP phase in the mixed phase~\cite{Bjorken}.
Here $r_d=\frac{g_Q}{g_H}$ is the ratio of the degrees of freedom in the
two phases and $\tau_f$ is the time when the temperature reaches $T_c$.
The evolution equation is then scaled accordingly for $\tau > \tau_f$.
We take the critical temperature $T_c$ to be $160\,\MeV$.

\begin{figure}
\centerline{\psfig{file=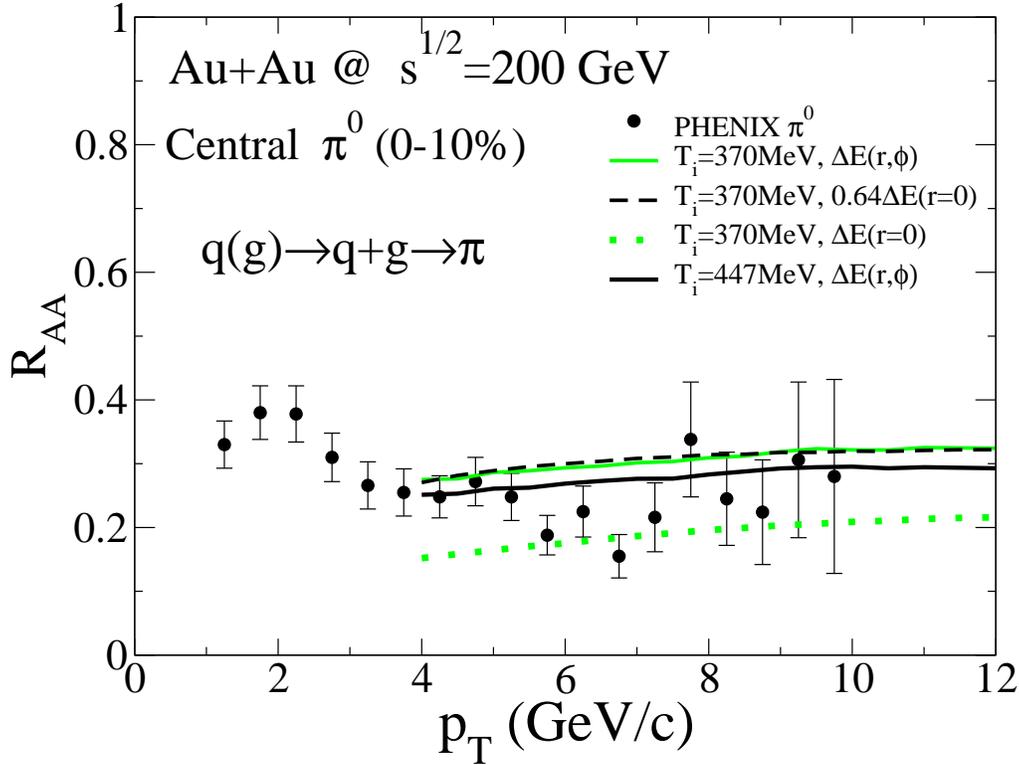,width=0.6\tw,angle=-90}}
\caption{(Color online) Nuclear modification factor for pions at
RHIC. Data points are from
PHENIX~\cite{phenix2}.  The solid lines show the full
calculation of the spatial distribution of jets in the plane
$z=0$ for two initial temperatures. 
The dotted lines assume that all jets are created at the center and the
dashed lines assume the same approximation but with a reduced energy loss.}
\label{Raa}
\end{figure}

The result of our calculation for RHIC energy
is summarized in Fig.~\ref{Raa} together with
the PHENIX data~\cite{phenix2}.
To cover the uncertainties in the initial conditions, we consider two
different sets,
one at a relatively high temperature of $T_i=447\, \MeV$
and a relatively short initial time of $\tau_i=0.147\, \fm/c$
taken from Refs.~\cite{prlphoton,prcdilep}, and
one at a lower temperature $T_i=370\, \MeV$ and somewhat later time
for the hydrodynamic evolution of $\tau_i=0.26\, \fm/c$ taken from
Ref.~\cite{simon}.  Those two sets correspond to ${dN}/{dy}|_{y=0}\sim 1260$,
estimated for central collisions in Ref.~\cite{prcdilep}.  In deriving
these results, we have used $\alpha_{\rm s}=0.3$.  If the value is
larger then energy loss will be greater, so $R_{AA}$ will be smaller; if
it is smaller, then $R_{AA}$ increases.  Besides this dependence on
$\alpha_{\rm s}$, our results rely on no free parameters.

The solid lines in the figure are full
calculations with the initial spatial distribution
$\calP(\bfr_\perp)$ given in Eq.~(\ref{Pr_dep}).
For comparison, we also display two more sets of calculations.
The dotted line is calculated with the jets originating only from the
center of the disk ($\bfr_\perp = 0$).
Comparing the dotted line with the solid line, it is apparent that the
absolute magnitude of the $R_{AA}$ depends very much on the density profile
of the nucleus.  The dashed line is calculated again with the jets from
the center but with the energy loss rates reduced by a factor of 0.64.  One
may say that the average path of a jet has the length of about
$0.64\times R_A$.

In Fig.~\ref{Raa}, one can see that
both $T_i = 447\,\MeV$ and $T_i = 370\,\MeV$
describe the real data reasonably well.
This is somewhat surprising.
Since the density of thermal particle is proportional to $T^3$, the density
at $447\,\MeV$ is about 1.8 times the density at $370\,\MeV$.  Yet the
energy loss does not reflect such a big difference.
The reason is because the energy loss depends mostly on
the duration of evolution.
In a Bjorken expansion, the initial time $\tau_i$ and the temperature are
related by
\be
T_i^3\tau_i=\frac{\pi^2}{\zeta (3)g_Q}\frac{1}{\pi R_{\perp}^2}\frac{dN}{dy}
\, .
\ee
Hence once $dN/dy$ is fixed, the time evolution of the temperature follows a
common curve regardless of the initial temperature
(c.f.~Eq.~(\ref{Tevolution})).  The only difference between the
higher and the lower temperature cases is that the time evolution starts
earlier for the higher temperature case.  From the moment $\tau$ passes
$\tau_i$ for the lower temperature, the evolution of the two systems is
identical.  Therefore, if energy loss at the beginning of the evolution is
small compared to the later time energy loss, the amount of energy loss
depends mostly on the duration of the QGP phase $\Delta \tau=\tau_f -
\tau_i$.  Since $\tau_f \gg \tau_i$, the duration of the QGP phase is
approximately the same for high temperatures.
We have verified that the energy loss between the times corresponding to
$T_i=1000\,\MeV$, $T_i=447\,\MeV$ and $T_i=370\,\MeV$ are at most about
10\,\%.

\begin{figure}
\centerline{\psfig{file=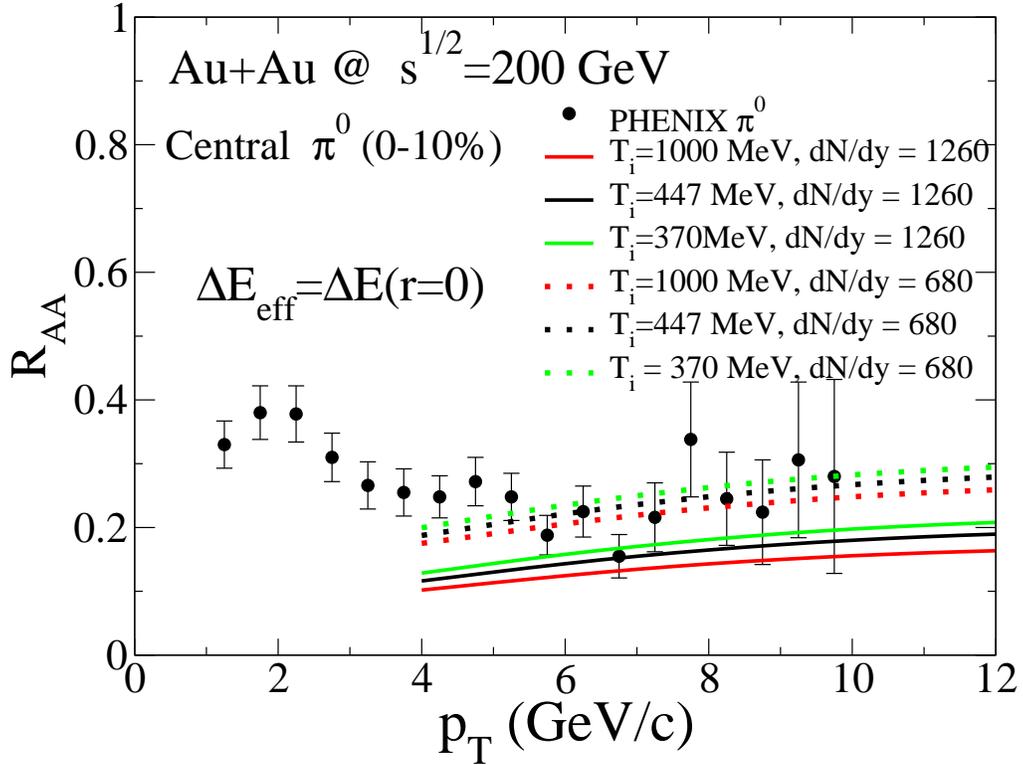,width=0.6\tw,angle=-90}}
\caption{(Color online) Nuclear modification factor for pions at
RHIC, taking all jets to originate at the center of the nucleus.
Data points are from
PHENIX~\cite{phenix2}. The solid lines are for initial conditions
which lead to a particle multiplicity of
$dN/dy=1260$ while the dotted lines are for initial conditions leading
to $dN/dy=680$.
For each set of lines, the initial temperature is, from bottom to top,
1000, 447 and 370 MeV.}
\label{Raa_dn}
\end{figure}

Since the suppression is mainly controlled by the duration
of the evolution, the suppression should be sensitive to the particle rapidity
density $dN/dy$, which fixes the lifetime of the QGP.  We can see in
Fig.~\ref{Raa_dn} that it is indeed the case.  For simplicity, we assume here
that the jets are all created at the center of the system.  We see that
there is very little change in the suppression as the initial
temperature varies from $T_i=370$ MeV to $T_i=1000$ MeV.  However, the
suppression shows a strong dependence on $dN/dy$; going from $dN/dy=680$ to
$dN/dy=1260$ increases the suppression by a factor $\sim$ 1.5.

For the LHC, we use the initial temperature from Ref.~\cite{simon},
$T_i=$845 MeV,
giving $\tau_i=$ 0.088 fm/c for $dN/dy \sim$ 5625~\cite{prcdilep,kms}.
At this energy, jets could in principle be produced with energies as high
as $\sqrt{s}/2=2750$ GeV.
However, according to our discussion in Sec.~\ref{sec:qual}, the
contribution from a jet having more that twice the observed pion energy
should be sharply cut off by the steeply falling initial function.
Since our approach is limited to the observed $p_T \lsim 30\,\GeV$,
considering original energy of only up to about $100\,\GeV$ should be enough
for our purposes.  To be on the conservative side, we cut off the maximum
jet energy at $400\,\GeV$.

\begin{figure}[ht!]
\psfig{file=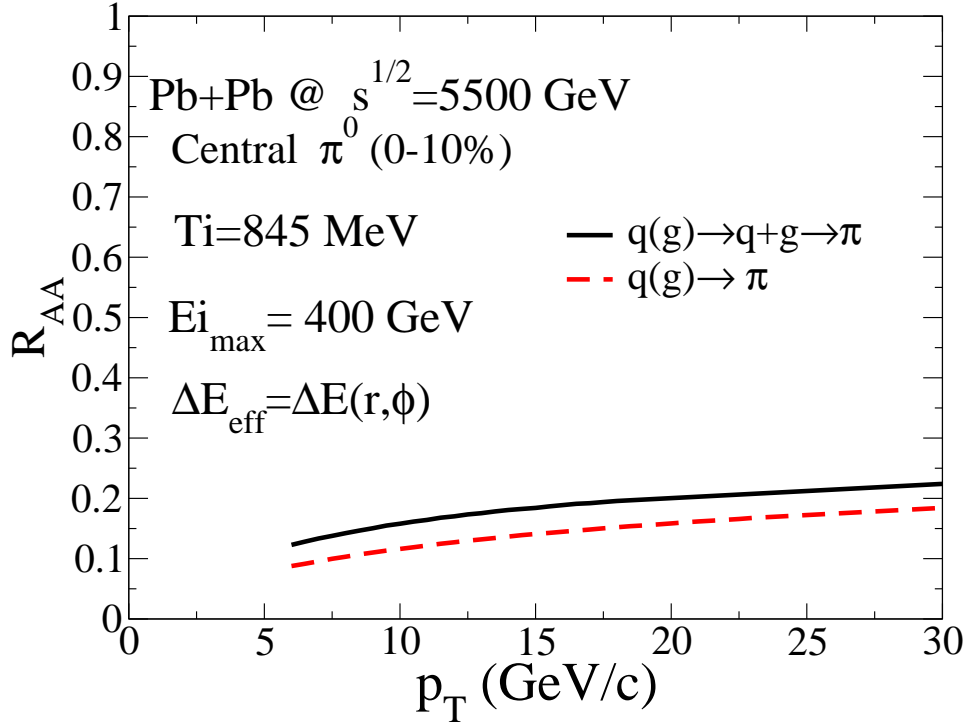,width=4in,angle=-90}
\caption{(Color online)
Nuclear modification factor of pions at the LHC.  The solid
line includes pions coming from bremsstrahlung secondaries emitted in
the thermal medium; the dashed line does not.
For both lines, the full spatial distribution from
Eq.~(\ref{Pr_dep}) has been used.}
\label{Raa_qtog}
\end{figure}

The calculated LHC nuclear modification factor is shown in
Fig.~\ref{Raa_qtog} with the full fragmentation function from
Eq.~(\ref{D_conv}).
We also show in the same figure the effect of not tracking the secondary
partons (gluons emitted by quarks and quarks emitted by gluons).
Comparison between the solid line and the dashed line shows that
ignoring the secondaries can make an overall difference of $30\%$.
This arises almost entirely from quark secondaries from gluon jets--that
is, from hard gluons which split, due to plasma interactions, into
$q\bar{q}$ pairs, which subsequently fragment.  This is important at the
LHC, but much less so at RHIC, because most hard jets at the LHC arise
from gluons, whereas a larger fraction at RHIC are from quarks.  The
gluons emitted by quarks are mostly soft; they also lose energy quickly
and fragment inefficiently.  At RHIC energies the error from dropping
secondaries is only $5\%$.

Comparing the RHIC and LHC results,
we see that the suppression due to
jet energy loss at RHIC is about a factor of 3 for a 10 GeV pion, while the
suppression at the LHC for
a pion at the same energy is about a factor of 6.
This difference arises because the medium at the LHC remains hot for
much longer than at RHIC.  The cooling is governed by the product
$T^3\tau \propto dN/dy$.  Hence, smaller $dN/dy$ implies faster cooling.
We have also calculated the impact of assuming a first order phase
transition. At RHIC, at the end of the QGP phase, a large number of jets
has already left the medium, but those still inside will suffer additional
suppression during the mixed phase, such that assuming a cross-over
between the QGP and the hadron gas phase will increase $R_{AA}$ by $\sim
20\%$.   At the LHC, at the critical point, a much larger fraction of jets are
out of the medium, such that $R_{AA}$ is quite insensitive to 
assumptions related to the transition.
Our initial conditions correspond to an initial time smaller than the
common value of $\tau_i$=0.6 fm/c used in hydrodynamic
calculations~\cite{hirano}. We have verified that going from $\tau_i$=.26
fm/c ($T_i=370$ MeV)  to 0.6 fm/c ($T_i$=280 MeV) increases $R_{AA}$ by
less than $20\%$.
In this section, the AMY formalism has been applied to successfully
reproduce the $\pi^0$ spectra at RHIC.
We will see in the following section, how the same formalism can be applied
to determine the production of real photons.

%%%%%%%%%%%%%%%%%%%%%%%%%%%%%%%%%%%%%%%%%%%%%%%%%%%%%%%%%%%%%%%%%%%%%%%%%%%%
\section{Photon Production}
\label{photon}
%%%%%%%%%%%%%%%%%%%%%%%%%%%%%%%%%%%%%%%%%%%%%%%%%%%%%%%%%%%%%%%%%%%%%%%%%%%%

\begin{figure}
\begin{center}
\psfig{file=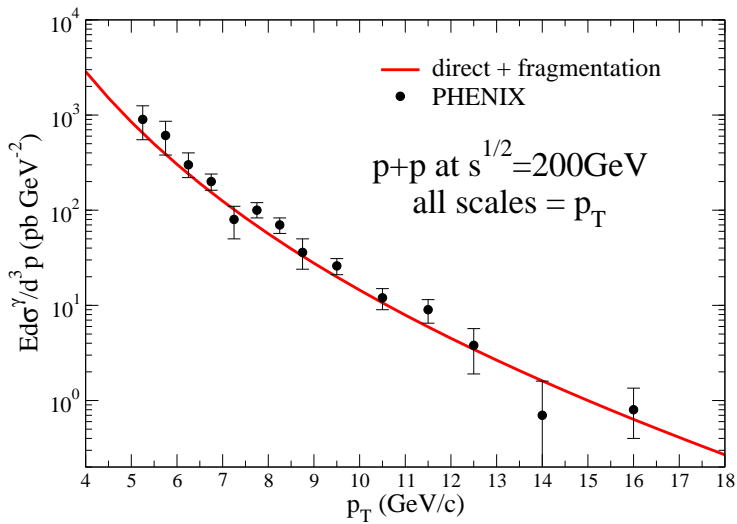,width=8.5cm,angle=-90}
\end{center}
\caption{(Color online) Prompt photons produced in $pp$ collision at
RHIC.  Data points are
from PHENIX.  The solid line is calculated with Eq.~(\ref{pp_prompt}). }
\label{gamma_pp}
\end{figure}

The hard photons produced in nucleon-nucleon collisions can be divided into 
three categories:
direct photons, fragmentation photons, and background photons.  Direct
photons are those produced by Compton scattering and annihilation of two
incoming partons.  Fragmentation photons are those produced by
bremsstrahlung emitted from final state partons.  Background photons are
those produced by the decay of hadrons subsequent to the collision,
primarily from $\pi^0 \rightarrow \gamma \gamma$ decay.
The `prompt photons' are those coming from direct production and the
fragmentation process.  The expression for prompt photon production is
\be
\label{pp_prompt}
E_{\gamma}\frac{d\sigma}{d^3p_{\gamma}}=\sum_{a,b}\int dx_a dx_b
g(x_a,Q)g(x_b,Q) &&\Big\{K_{\gamma}(p_T)\frac{d\sigma^{a+b\rightarrow
\gamma+d}}{dt}\frac{2x_ax_b}{\pi
(2x_a-2\frac{p_T}{\sqrt{s}}e^y)} 
\delta \left(x_b - \frac{x_a p_T e^{-y}}{x_a \sqrt{s} - p_T
e^y}\right)
\nonumber \\
&&+K_{brem}(p_T)\frac{d\sigma^{a+b\rightarrow
c+d}}{dt}\frac{1}{\pi z} D_{\gamma/c}(z, Q)\Big\} \, .
\ee
$K_{\gamma}$ and $K_{brem}$ are correction factors to take into account NLO effects; we
evaluate them using the numerical program from Aurenche 
{\it et al.}~\cite{aurenche}, 
obtaining $K_{\gamma}$(10 GeV)$\sim$ 1.5 for RHIC and LHC and $K_{brem}$(10 GeV)$\sim$ 1.8 at RHIC and 1.4 at LHC.  All scales (renormalization, factorization and fragmentation) have been set equal to the photon transverse momentum $p_T$.
 The photon fragmentation function $D_{\gamma/c}$ is
extracted without medium effects in $e^-+e^+$
collisions~\cite{photonfrag}.
The validity of this expression
for $pp$ collisions at $\sqrt{s}=$ 200 GeV
is shown in Fig.\ref{gamma_pp} with data from PHENIX
(Klaus Reygers, Hard Probes 2004).
It appears clear that the baseline mechanism of high $p_T$ photon production
in nucleon-nucleon collisions is under quantitative control.

In $AA$ collisions, there is an additional source of high $p_T$ photons: the
medium contribution.  This contribution includes the direct conversion of
a high energy parton to a high energy photon by annihilation
with a thermal parton,
in-medium bremsstrahlung from a jet, and
thermal production of photons.

\subsection{Jet-Photon Conversion}

In Ref.~\cite{prlphoton}, the conversion of a leading parton
to a photon in the plasma was found to be an important process.
This happens when a jet crossing the hot medium
undergoes an annihilation ($q+\bar{q}\rightarrow g+\gamma$)
or a Compton ($g+q\rightarrow q+\gamma$) process with a thermal parton.
The related production rate of photons is given by~\cite{prlphoton,wong}
\begin{equation}
\label{photonrate}
\frac{dR}{dyd^2p_T}
=\sum_f\left(\frac{e_f}{e}\right)^2
\frac{T^2\alpha\alpha_s}{8\pi^2}
[f_q(\overrightarrow{p_{\gamma}})+f_{\bar{q}}(\overrightarrow{p_{\gamma}})]
\left[2\ln \left(\frac{4E_{\gamma}T}{m^2}\right)
	-C_{\rm ann}-C_{\rm Com}\right]\,,
\end{equation}
where $T$ is the temperature, $C_{\rm ann}$=1.916, 
and $C_{\rm Com}$=0.416.  It has been
shown in Ref.~\cite{kapusta,baier_gamma}, that the 
infrared dependence on the quark mass should be replaced by a dependence
on the in-medium thermal quark mass,
$m^2=2m^2_{th}=4\pi\alpha_sT^2/3$.
The phase-space distribution function of the incoming particles is defined as
\begin{equation}
g_i \int \frac{d^3x d^3p}{(2\pi)^3} f(x,p) = N_i \, ,
\end{equation}
where $N_i$ is the number of particles $i$ and $g_i$ is the spin-color
degeneracy.
The phase-space distribution function for an incoming jet, assuming a Bjorken
$\eta-y$ correlation~\cite{lin}, is
\be
f_{jet}(\overrightarrow{x},\overrightarrow{p},t_0)&=&\frac{(2\pi)^3
\calP(\bfr_\perp)}{g_q\tau p_T}
\frac{dN^{jet}}{dyd^2p_T}\delta(\eta-y)\nonumber \\ &=&\frac{(2\pi)^3
\calP(\bfr_\perp)t_0}{g_q\sqrt{t_0^2-z^2}}\frac{p_T}{E^2}
\frac{dN^{jet}}{dyd^2p_T}\delta(z-v_zt_0)\, ,
\ee
where $\eta$ is the space-time rapidity and $t_0$ is the formation time
of the jets.  As before, the jets are taken to be massless and we suppose that
energy loss of jets in the plasma does not change their direction. With this
latter approximation, $f_{jet}$ can be factorized into a position space
and a momentum space part:
\be
f_{jet}(\overrightarrow{x},\overrightarrow{p},t)&=&
\chi(\overrightarrow{x},t)
\frac{1}{E^2}\frac{dN^{jet}(E,t)}{dE}\nonumber \\
&=&\chi(\overrightarrow{x}-\hat{t}
\frac{\overrightarrow{p}}{\left|\overrightarrow{p}\right|},t_0)
\frac{1}{E^2}\frac{dN^{jet}(E,t)}{dE} \, ,
\ee
where $\hat{t}=t-t_0$ is the propagation time of the jet.
In the high energy limit, the cross-sections $\sigma_{q\bar{q}\rightarrow
g\gamma}$ and $\sigma_{qg\rightarrow q\gamma}$ are dominated by direct exchange
between the quark and the photons.  Since
we are interested in photons produced at mid-rapidity ($y=0$),
we only need to
consider quark and anti-quark jets produced at mid-rapidity.  This gives
\be
\label{fjet_t}
f_{jet}(\overrightarrow{x},\overrightarrow{p},t)\big|_{y=0}=\frac{(2\pi)^3
\calP(\left|\overrightarrow{w_r}\right|)t_0}{g_q
\sqrt{t_0^2-z^2}}\frac{1}{p_T}
\frac{dN^{q\bar{q}}}{dyd^2p_T}(p_T,t)\delta(z) \, ,
\ee
where $\overrightarrow{w_r}$ is, in the plane $z=0$, the initial radial
position of the jet,
\be
\left|\overrightarrow{w_r}\right|
=\left(\overrightarrow{x}-\hat{t}
\frac{\overrightarrow{p}}{\left|\overrightarrow{p}\right|}\right)
\cdot\hat{r}
=\sqrt{(r\cos\phi-t)^2+r^2 \sin^2\phi}\quad  \mbox{for $t_0 \sim0$} \, ,
\ee
and $\phi$ is the angle in the plane $z=0$ between the direction of the photon
and
the position where this photon has been produced.  The AMY formalism is
introduced here to calculate the evolution of
$\frac{dN^{q\bar{q}}}{dyd^2p_T}(p_T,t)$.  The output of Eq.~(\ref{eq:Fokker})
is $\Pq(E,t)$, where
\be
\Pq(E,t)=\frac{dN^{q\bar{q}}}{dE}(E,t)\propto
p_T\frac{dN^{q\bar{q}}}{dyd^2p_T}(p_T,t)\, .
\ee

The initial distributions
$\Pq(E,t_0)$ and $\Pg(E,t_0)$ are fixed by the initial jet
distribution at mid-rapidity
as parametrized in Ref.~\cite{prlphoton,prcdilep}.
The total photon spectrum is given by a full space-time integration:
\be
\label{photon_yield}
\frac{dN^{\gamma}}{dyd^2p_T}&=&\int d\tau \tau\int rdr\int d\phi\int d\eta
\frac{dR}{dyd^2p_T}(E_{\gamma}=p_T \cosh(y-\eta))\nonumber \\ &=&\int dt \int
rdr\int
d\phi\int dz
\frac{dR}{dyd^2p_T}(E_{\gamma}=p_T \cosh(y_0))\, .
\ee
The production rate is calculated in the local frame where the temperature is
defined where the photon rapidity becomes $y_0 = y - \eta$.

High-$p_T$ photons 
are emitted 
preferentially early during the QGP phase, when the temperature is at its
highest point. Indeed, explicit hydrodynamic calculations show that the
nuclear space-time geometry smoothly evolve from 1-D to 3-D~\cite{kolb}. 
By the time the system reaches the temperature corresponding
to the mixed phase in a first-order phase transition, the system is still very
much 1-D~\cite{kolb}. For such a geometry, specific calculations~\cite{flow} 
 suggest that the flow effect on 
photons and dileptons from the QGP is not large at RHIC and LHC for
 $p_T > 2$ GeV.  Assuming again a 1-D expansion, we get the production 
of photons
from jet-medium interactions 
from Eqs.~(\ref{photonrate}),~(\ref{fjet_t}) and~(\ref{photon_yield}),
\be
\frac{dN^{\gamma}_{jet-th}}{dyd^2p_T}\Big|_{y=0}&=&2\int dt\int_0^{R_{\perp}}
rdr\int_0^{\pi} d\phi
\frac{(2\pi)^3 \calP(\left|\overrightarrow{w_r}\right|)}{g_q}\frac{1}{p_T}
\frac{dN^{q\bar{q}}}{dyd^2p_T}(p_T,t)\nonumber \\ &
&\sum_f\left(\frac{e_f}{e}\right)^2
\frac{T^2\alpha\alpha_s}{8\pi^2}
\left[2\ln\left(\frac{3p_t}{\pi\alpha_sT}\right)
-C_{\rm ann}-C_{\rm Com}\right] \, ,
\ee
where the temperature $T$ evolves according to Eq.~(\ref{Tevolution}).  The
$\phi$ integration can be done:
\begin{eqnarray}
\label{gamma_rt}
\int_0^{\pi} d\phi
\calP(\left|\overrightarrow{w_r}\right|)=\gamma(r,t)
=
\left\{
\begin{array}{cc}
0 &
\displaystyle
 r^2+t^2-2tr > R_{\perp}^2 \\
\displaystyle
\frac{2}{R_\perp^2}
\left(1-\frac{r^2+t^2}{R_{\perp}^2}\right)
& r^2+t^2+2tr < R_{\perp}^2 \\
\displaystyle
\frac{2u_0}{\pi
R_{\perp}^2}\left(1-\frac{r^2+t^2}{R_{\perp}^2}\right)+\frac{4tr}{\pi
R_{\perp}^4}\sin (u_0)
& \mbox{otherwise}\,,
\end{array}
\right.
\end{eqnarray}
where
\begin{equation}
u_0=\arccos \left(\frac{r^2+t^2-R_{\perp}^2}{2tr}\right)\,.
\end{equation}
Then, the final expression for the jet-photon production becomes
\be
\frac{dN^{\gamma}_{jet-th}}{dyd^2p_T}\Big|_{y=0}=& & 2\int dt\int_0^{R_{\perp}}
rdr
\frac{(2\pi)^3}{g_q}\frac{1}{p_T}
\frac{dN^{q\bar{q}}}{dyd^2p_T}(p_T,t)
\gamma(r, t)
\nonumber \\ &
&\sum_f\left(\frac{e_f}{e}\right)^2
\frac{T^2\alpha\alpha_s}{8\pi^2}
\left[2\ln \left(\frac{3p_t}{\pi\alpha_sT}\right)
-C_{ann}-C_{Com}\right]\nonumber \\ &= & \int dt
\frac{dN^{\gamma}_{jet-th}}{dt dyd^2p_T}\Big|_{y=0} 
\ee

\begin{figure}
\psfig{file=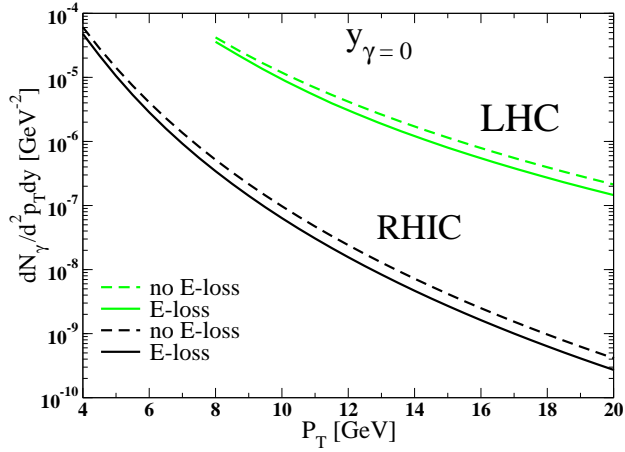,width=6.5cm,angle=-90}
\caption{(Color online) Direct production of photons by jets in the plasma for
Au-Au at RHIC and Pb-Pb at the LHC.  For the solid lines, jet energy
loss is included; for the dashed lines it is neglected.
The initial temperature is $T_i=370$ MeV at RHIC and $T_i$=845
MeV at the LHC.}
\label{rhicjetth}
\end{figure}

As before, we assume a first order phase transition
beginning at the time $\tau_f$
and ending at $\tau_H = r_d \tau_f$~\cite{Bjorken}.
After $\tau_f$  we scale the production
rate by $f_{QGP}$ (Eq.~(\ref{fmix})), such that
\be
\frac{dN^{\gamma}_{jet-th}}{dyd^2p_T}\Big|_{y=0}=\int_{\tau_i}^{\tau_f}
dt \frac{dN^{\gamma}_{jet-th}}{dt
dyd^2p_T}\Big|_{y=0}+\int_{\tau_f}^{\tau_H} dt f_{QGP}(t)
\frac{dN^{\gamma}_{jet-th}}{dt dyd^2p_T}\Big|_{y=0}
\ee
The first term and second term correspond respectively to photons
produced during the pure QGP and the mixed phase.  Our results for RHIC
and LHC, with and
without energy loss are shown in Fig.~\ref{rhicjetth}. 
Here also, $\alpha_s=0.3$. We see that higher
energy photons are more sensitive to jet energy loss:  photons at 4 GeV are
suppressed by a factor 1.3 at RHIC, while 15 GeV photons are suppressed by a
factor 1.6 due to jet energy loss.
We find approximately the same suppression for the LHC.
This suppression is much smaller that the one observed from $R_{AA}$ in the
previous section.  This is because the photons can be produced at any
point in the hard parton's propagation through the medium, while the jet
energy loss depends on the final parton energy.  Hence, some of the
photon rate arises from before, rather than after, the jet has lost much
energy.

\subsection{Bremsstrahlung Photons}

Hard partons in the medium can also produce photons by bremsstrahlung
when they scatter in the medium.  The photon
bremsstrahlung rate $\frac{d\Gamma^{q\rightarrow q\gamma}(p,k)}{dk dt}$
follows the same expression as $\frac{d\Gamma^{q\rightarrow qg}(p,k)}{dk dt}$
in Eq.~(\ref{eq:dGamma}), but with $C_s\rightarrow \frac{e_f^2}{e^2}$,
$C_{\rm A}
\rightarrow 0$, $m_{\gamma}=0$ and $g_s^2\rightarrow e^2$.  The photon
bremsstrahlung distribution  is given by
\be
\frac{dP_{\gamma}(p,t)}{dt}=\int dk \: \Pq(p+k) \frac{d\Gamma^{q\rightarrow
q\gamma}(p{+}k,p)}{dk dt} \, .
\ee
It is assumed here that the photon production rate is so low that the quark
plus anti-quark distribution $\Pq$ will be unchanged by the photon emission.
The photon distribution is finally convolved with the initial spatial
distribution of jets to get the final spectrum of bremsstrahlung photons,
\be
\frac{dN_{jet-br}}{dyd^2p_T}
\Big|_{y=0}=\frac{1}{p_T\Omega(y=0)}
\frac{dN_{jet-br}}{dp_T}
=\frac{1}{p_T}\int d^2 r_\perp\,
\calP(\bfr_\perp)P_{\gamma}(p_T,d)\, ,
\ee
where $r_\perp$ is the position where the jet has been created, and
$d=d(r_\perp)$ is the distance crossed by the jet in the plasma. The factor
$\Omega(y=0)$ corresponds to a $d\phi$ and a $dy$ integration around the plane
$y=0$.  This factor can be absorbed in the definition of the initial
distribution $\Pq(p_T,t=0)$.

Numerically, bremsstrahlung photons turn out to be subdominant to
jet-photon conversion.  This is because, while the rate at which such
photons are produced is larger, they typically carry only a fraction of
the jet's energy, while jet-photon conversion predominantly produces
photons with the complete energy of the jet (hard parton).  When folded
against a steeply falling spectrum of jets, the process which produces
the highest energy photons will dominate the final spectrum.

Recently, Zakharov has also considered bremsstrahlung emission of
photons from jets~\cite{zakharov2}.  His work accounts for finite-size
effects in the high energy limit; in this respect it is more complete
than our work here.  However, our energy range of
interest for this study, as discussed at the end of
section~\ref{lab:formalism}, is below the factorization scale, so finite
size effects are small for the energies we consider.

\subsection{Thermal Photons}
The thermal-thermal contribution comes from the photons produced by
two scattering thermal particles.
The Compton and annihilation rate has been calculated in the previous
literature \cite{kapusta,baier_gamma}, and we use those results here.
There are also leading order bremsstrahlung and inelastic pair
annihilation contributions \cite{guy3}; we use the parameterization for
those rates presented there.

\subsection{Non-thermal contributions}

The expression for prompt photons produced in $AA$ collisions is
\be
\frac{dN_{\gamma
-prompt}}{dyd^2p_T}=E_{\gamma}\frac{d\sigma}{d^3p_{\gamma}}
\frac{\ave{N_{\rm coll}}}{\sigma_{in}} \, ,
\ee
where we take the values $\ave{N_{\rm coll}}=975$, $\sigma_{in}=40$ mb for
RHIC~\cite{jeon} and $\ave{N_{\rm coll}}=1670$, $\sigma_{in}=72$ mb for
the LHC~\cite{yellow}.
$E_{\gamma}\frac{d\sigma}{d^3p_{\gamma}}$ is taken from Eq.~(\ref{pp_prompt})
but with the photon fragmentation function  accounting for the jet
energy loss.

We will assume, as we did for pion production, that photon production
via fragmentation of a jet occurs after the jet parton leaves the QGP.
Therefore, the photon fragmentation function
including the full spatial distribution and the secondary jets, is
given by Eq.~(\ref{D_conv}), with the substitution $\pi^0\rightarrow
\gamma$.

The preequilibrium contribution of photons, corresponding to photons emitted
after the transit time of the two nuclei but before thermalization time, is
not explicitly included in this work.  However, a rough estimate might be had
 by choosing a smaller formation time, as we do.  The modeling of those 
contributions is accessible by the parton cascade model~\cite{cascade}. 
Finally, in order to have a complete photon description, we have also
calculated the background production, which mainly comes from the decay
$\pi^0\rightarrow \gamma\gamma$.  This is given by~\cite{cahn}
\be
\label{back_pi}
\frac{dN_{\gamma
-BG}}{dyd^2p_T}=\int dy^{\pi^0} d^2p_T^{\pi^0}
\frac{dN^{\pi^0}}{dy^{\pi^0} d^2p_T^{\pi^0}}
\frac{dP(\bfp_{\pi^0}
\rightarrow\bfp_{\gamma})}{dy d^2p_T} \, .
\ee
All the previous procedures for jet energy
loss, initial spatial distribution and the effect of secondary jets are
included in the calculation of the pion spectrum $\frac{dN^{\pi^0}}{dy^{\pi^0}
d^2p_T^{\pi^0}}$.
In the pion center of mass frame, the photon distribution is given by
\begin{equation}
\frac{dP({\bfp_{\pi^0}}
\rightarrow{\bfp_{\gamma}})}{dy d^2p_T}
=\frac{\delta(E^{\gamma}_{cm}-\frac{m_{\pi^0}}{2})}{2\pi E^{\gamma}_{cm}}
\end{equation}
where
\begin{equation}
E^{\gamma}_{cm}=p_T\cosh y
\sqrt{\sin^2\theta
+\frac{(E_{\pi^0}
\cos\theta-|{\bfp_{\pi^0}}|)^2}{m_{\pi^0}^2}}
\end{equation}
and $\theta$ is the angle between ${\bfp_{\pi^0}}$ and
${\bfp_{\gamma}}$. With the $\eta$ branching ratio
$\Gamma^{\eta\rightarrow\gamma\gamma}/\Gamma^{\eta}\sim 40\%$~\cite{pdg2004}
and its relative yield $N^{\eta}/N^{\pi^0}\sim$ 0.5~\cite{albrecht}, we simply
multiply Eq.~(\ref{back_pi}) by a factor of
$1.2$ to include the $\eta$ contribution.

\subsection{Results}

\begin{figure}[ht!]
\psfig{file=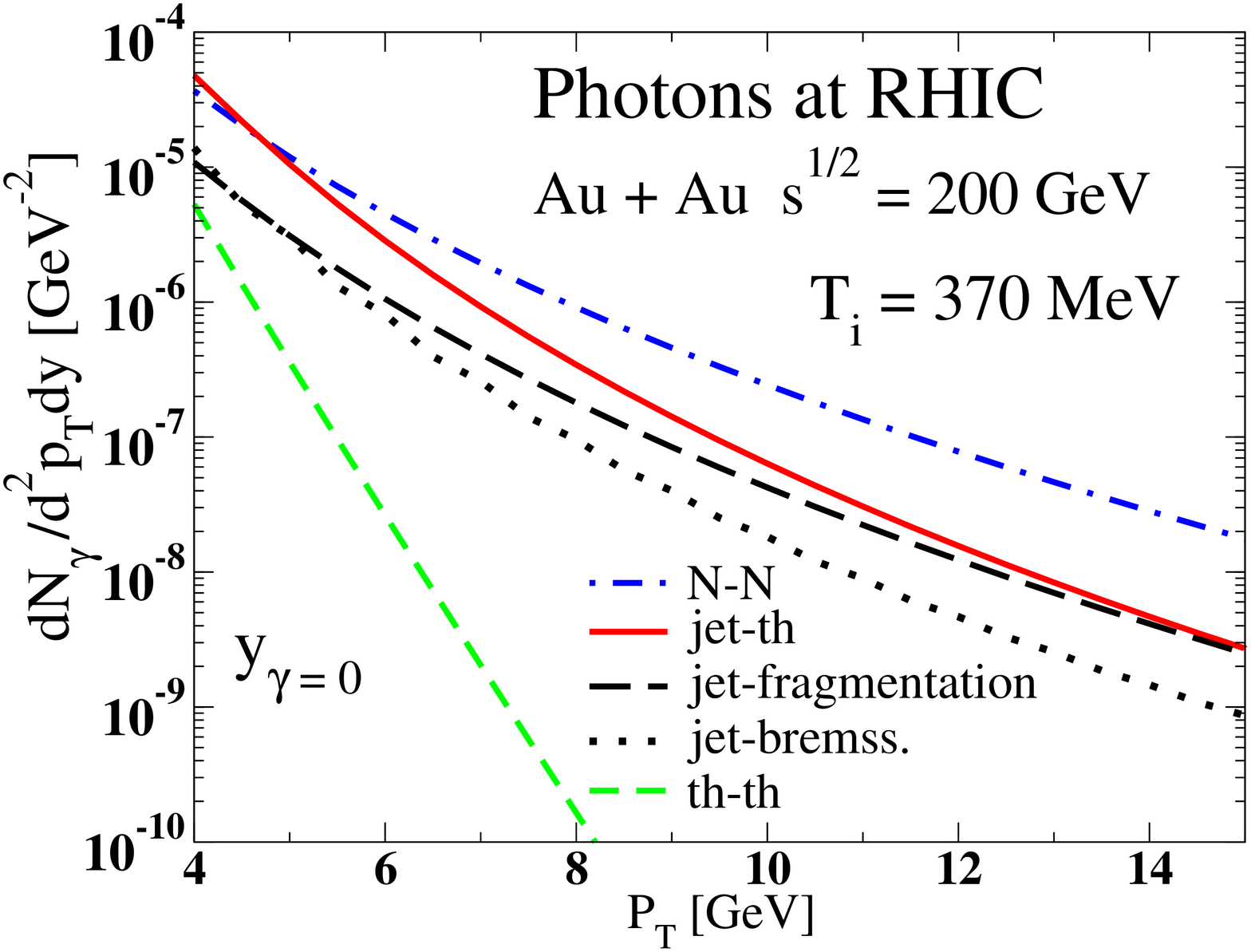,width=6.cm,angle=-90}
\psfig{file=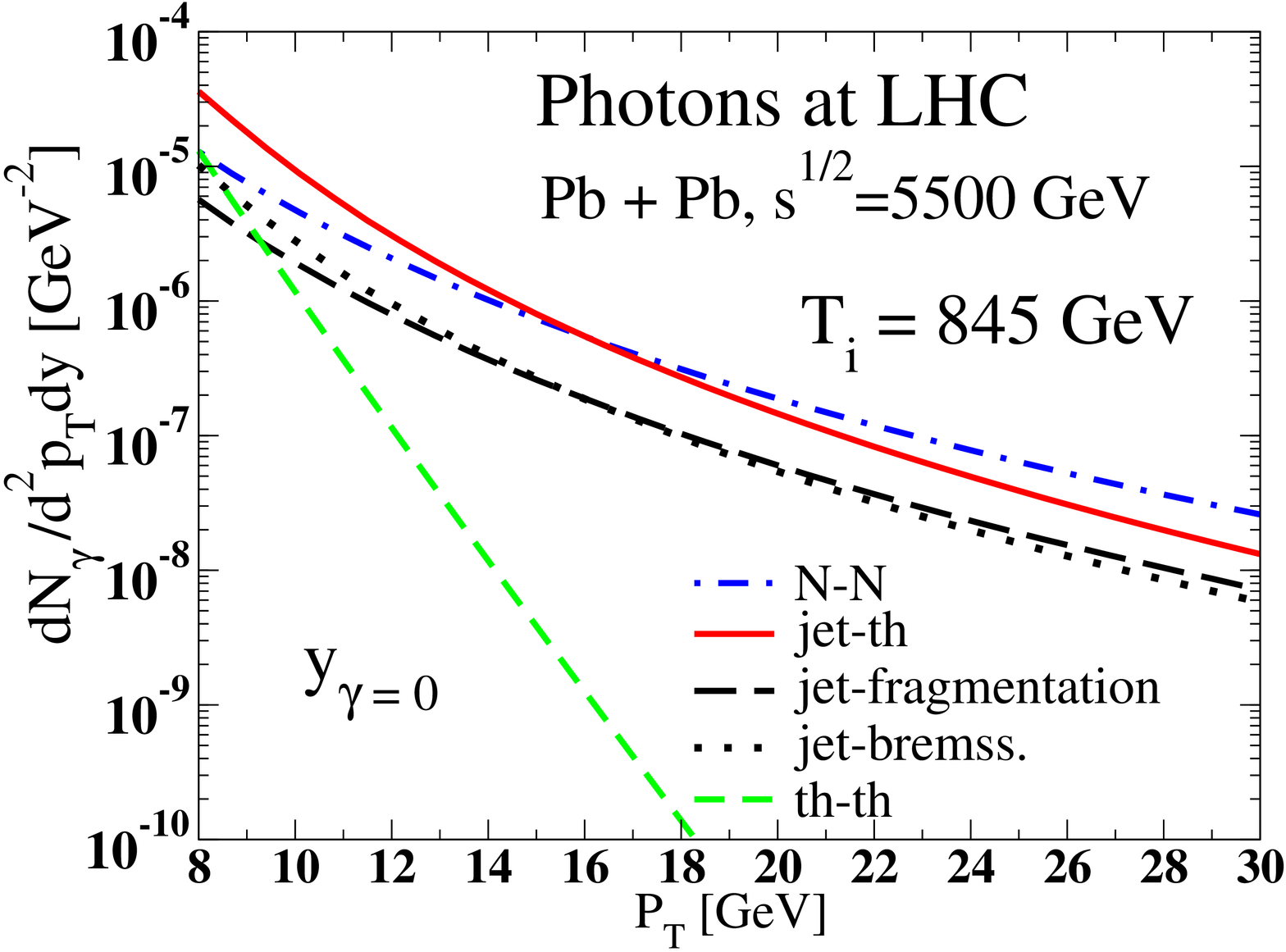,width=6.cm,angle=-90}
\caption{(Color online) Contributing sources of high-$p_T$ photons at
mid-rapidity in central Au-Au
collisions at RHIC (left panel) and Pb-Pb collisions at the LHC (right
panel). Solid line: jet-photon conversion in the plasma; dotted
line: bremsstrahlung from jets in the plasma; short dashed line: thermal
induced production of photons; long dashed line: fragmentation of jets
outside the plasma; and dot-dashed line:  direct contribution from the
primordial hard scattering.}
\label{rhic_source}
\end{figure}

Each contribution of high $p_T$ photons, except the background, are
shown in
Fig.~\ref{rhic_source} for central collisions at RHIC and the LHC.  The
energy loss is included in all processes involving jets.   Prompt
photons have been split into the direct component (N-N) and the
fragmentation component.  All those processes, except the
contribution from jet-medium bremsstrahlung, have been presented in
Ref.~\cite{prlphoton} for the case of no energy loss.  For RHIC, as in
Ref.~\cite{prlphoton}, the high $p_T$ region is dominated by direct
photons.  However, in Ref.~\cite{prlphoton}, 
the jet-photon conversion was dominant below 6 GeV, while in our study, direct
photons dominate all the high $p_T$ spectrum.  A few factors explain this difference.
The jet energy loss is included here; the constants $C_{\rm ann}$ and
$C_{\rm Com}$ appearing in Eq.~(\ref{photonrate}) have been set equal to
1.916 in Ref.~\cite{prlphoton};
the $K_{jet}$ factor in the original publication is larger than ours: 
$K_{jet}=2.5$ is used 
for both
RHIC and the LHC while we use $K_{jet}=1.7$ for RHIC and 1.6 for LHC.
Finally, no
$K_{\gamma}$ factor has been used for the direct contribution in
Ref.~\cite{prlphoton}. It is however satisfying
that the inclusion of jet energy loss does not spoil the original premise:
jet-photon conversion is an important source of electromagnetic radiation.

At the LHC, our
result is dominated by direct photons for $p_T$ above 20 GeV, but there is a
window, below 14 GeV, where the jet-photon conversion in the plasma is the
dominant mechanism of photon production.  In Ref.~\cite{prlphoton},
however, the jet fragmentation (called bremsstrahlung in their study)
was the most important
process at the LHC, but jet suppression was not included.  Photon
production via 
jet bremsstrahlung in the plasma (dotted lines) turns out to be weak, but
non-negligible.  It is approximately a factor 3 below the jet-photon
conversion contribution.  Finally, the thermal induced photons (short dashed
lines) are far below all other contributions in intensity.

\begin{figure}[ht!]
\psfig{file=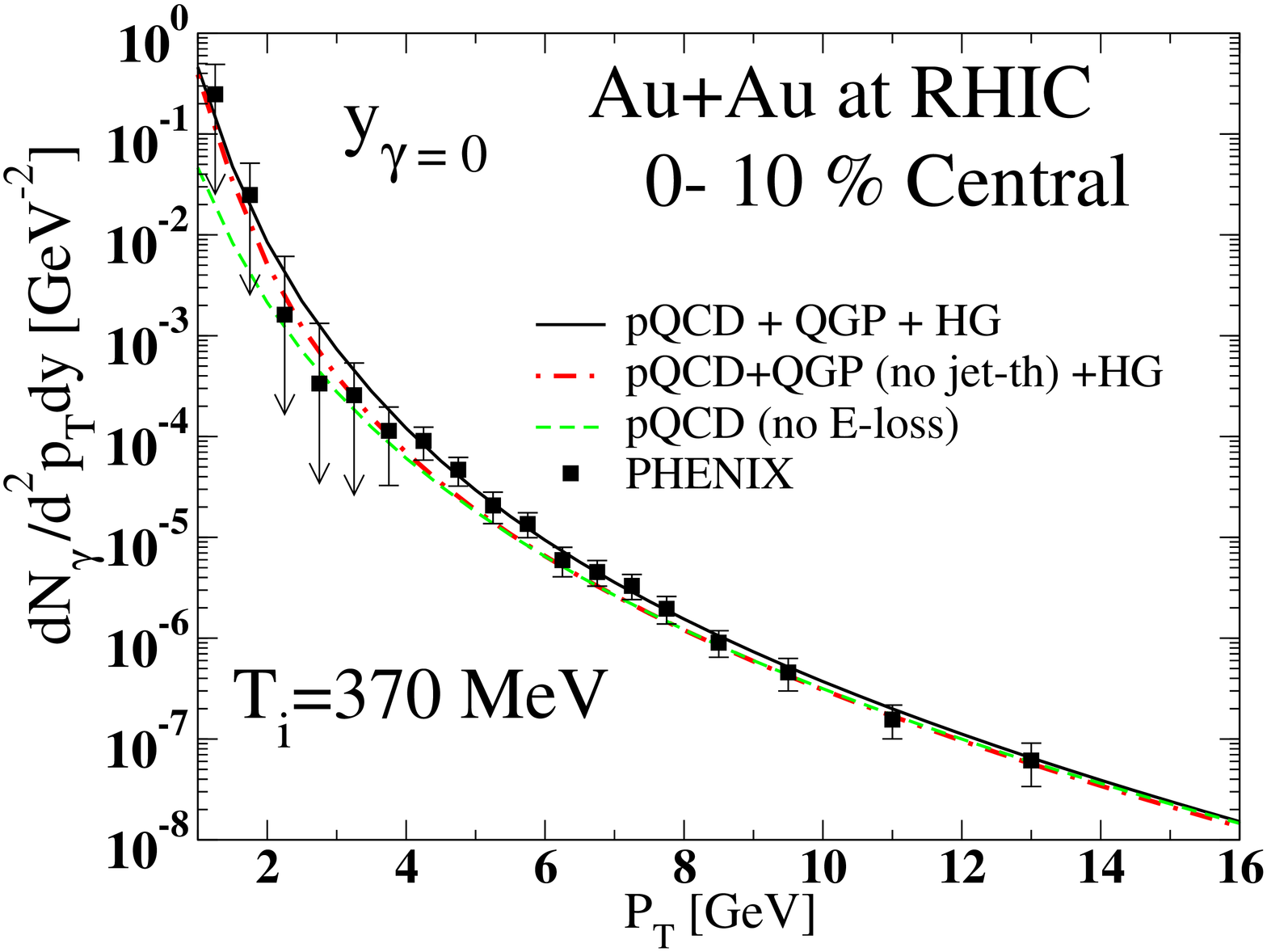,width=6.cm,angle=-90}
\psfig{file=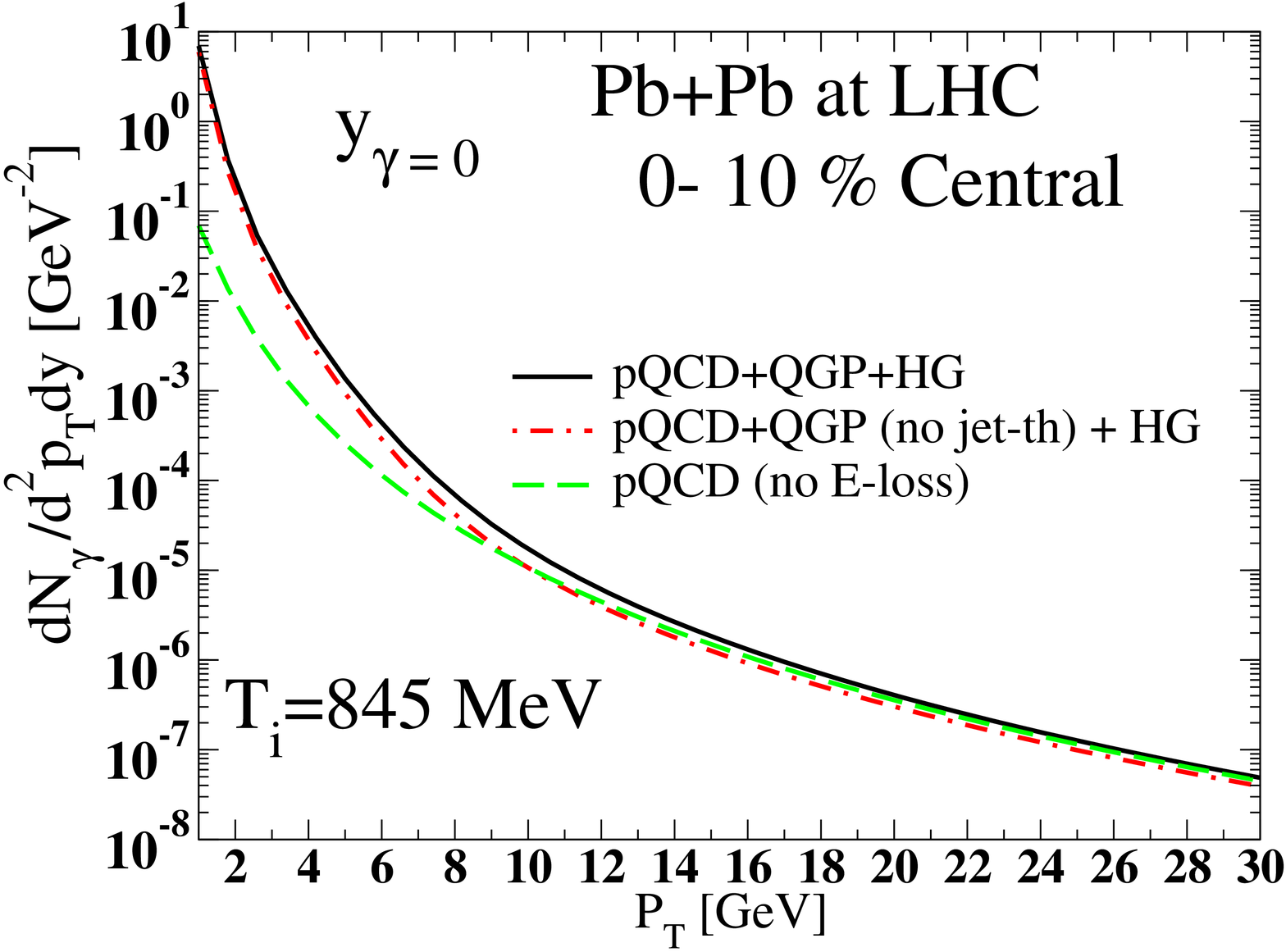,width=6.cm,angle=-90}
\caption{(Color online) Total production of photons in central Au-Au
collisions at RHIC and Pb-Pb at the LHC. The solid lines include all
process from Fig.~\ref{rhic_source} and the hadron gas
contribution~\cite{simon}, while the dot-dashed lines do not include the
jet-thermal contribution.  p-p collisions scaled to A-A are shown by the
dashed lines. Data at RHIC are from PHENIX~\cite{PHENIX_photon}. } 
\label{phrhic_sansth}
\end{figure}

New results for total photon production, after background subtraction, 
are now available~\cite{PHENIX_photon}.  Our calculations are compared to
experimental data on the left-hand side of
Fig.~\ref{phrhic_sansth}.  The solid line includes the prompt photon
contribution (pQCD), the  QGP (jet-th, th-th and jet-bremss.) and hadron
gas contribution, extracted from Ref.~\cite{simon}.  This latter
contribution becomes important only for $p_T <$ 1.5 GeV.  The initial
condition for the thermal phase corresponds to $T_i=370$ MeV and
$\tau_i$=0.26 fm/c. For a better 
comparison with data, we have extended our calculation down to $p_T$=1
GeV.  No cutoff has been applied on either jet-th or pQCD process.  NLO
calculations
are not very reliable is this region, but thermal induced
reactions and hadron gas contributions turn out to dominate here and NLO
results play a minor role. When the jet-photon conversion is not included
(dot-dashed line), the total photon production is reduced by up to 45
$\%$, around $p_T=3$ GeV, showing the importance of that process.  The
result expected from N-N collisions scaled to Au-Au is also shown (dashed
line).  The plasma contribution, specially the jet-thermal process, is
very important
for $p_T <$ 6 GeV.  However, large error bars prevent a strong claim about
the presence of a QGP.

\begin{figure}[ht!]
\psfig{file=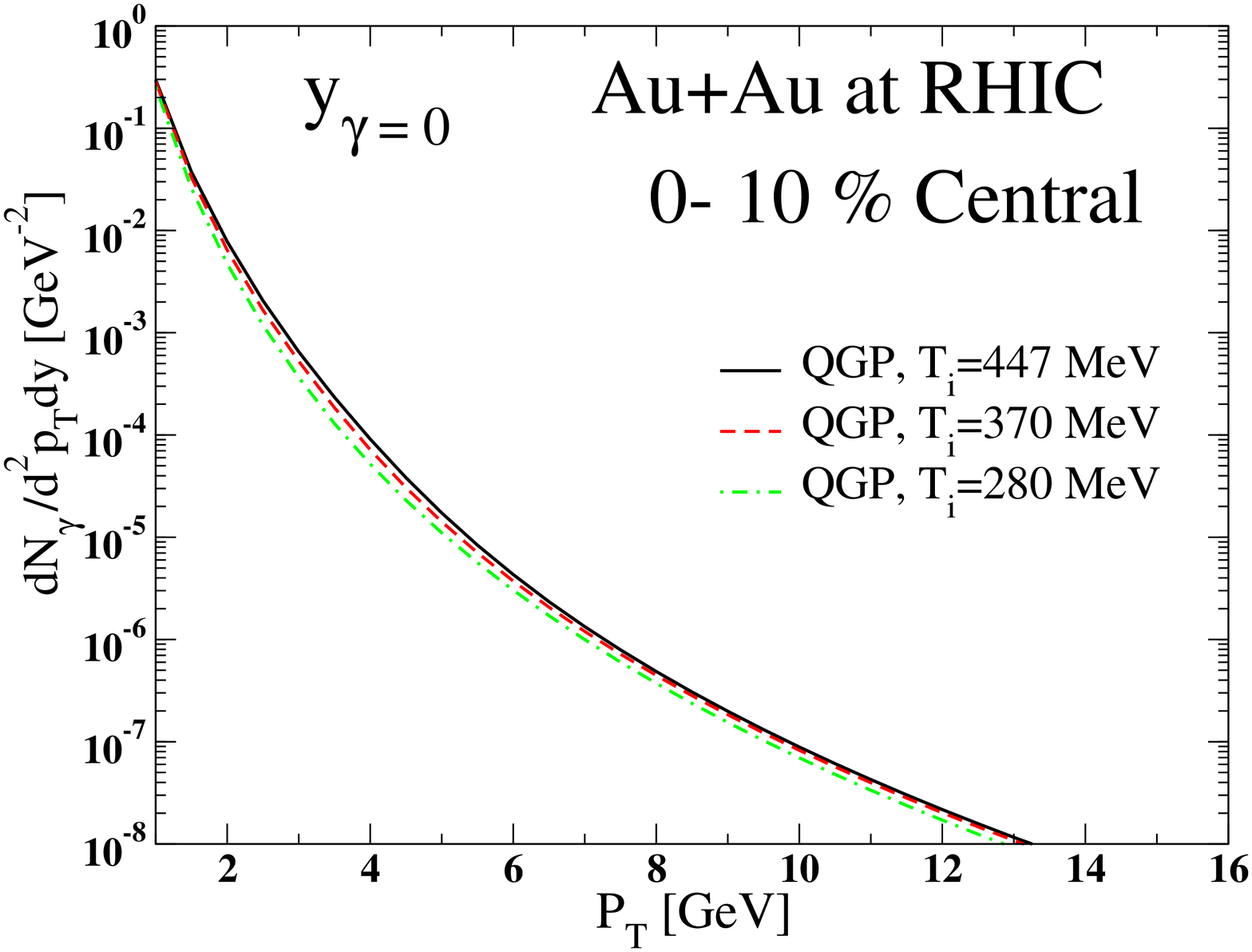,width=6.5cm,angle=-90}
\caption{(Color online) Production of photons during the QGP phase for
three QGP initial conditions: solid line, ($T_i=447$ MeV,$\tau_i$=0.147
fm/c); dashed line, ($T_i=370$ MeV,$\tau_i$=0.26 fm/c); and dot-dashed
line, ($T_i=280$ MeV,$\tau_i$=0.6 fm/c). } 
\label{Ti_effect}
\end{figure}

Fig.~\ref{Ti_effect} shows the QGP photons for three different initial
conditions: ($T_i=447$ MeV, $\tau_i$=0.147 fm/c),($T_i=370$ MeV,
$\tau_i$=0.26 fm/c) and 
($T_i=280$ MeV, $\tau_i$=0.6 fm/c).   As the high-$p_T$ photons are
produced
early in the collision, they may be affected by the initial conditions.
However, the high-$p_T$ region is dominated by jet-therm processes, which
are weakly sensitive to 
($T_i,\tau_i$), since the jet distribution function $f_{jet}$ has a
weak temperature dependence.  We see that going from $\tau_i$=0.6 fm/c
to $\tau_i$=0.26 fm/c increases the photon production in the QGP phase by
less that a factor 2; this additional contribution could be interpreted in
some sense as a preequilibrium contribution. 

Our prediction for the LHC is shown in the right-hand side of
Fig.~\ref{phrhic_sansth}. The signature of the QGP phase is
much stronger than at RHIC, increasing the photon yield, relatively to N-N
scaled to A-A collision, by one order of magnitude around $p_T=3$ GeV,
where the hadron gas contribution turns out to be negligible. 

\begin{figure}
\psfig{file=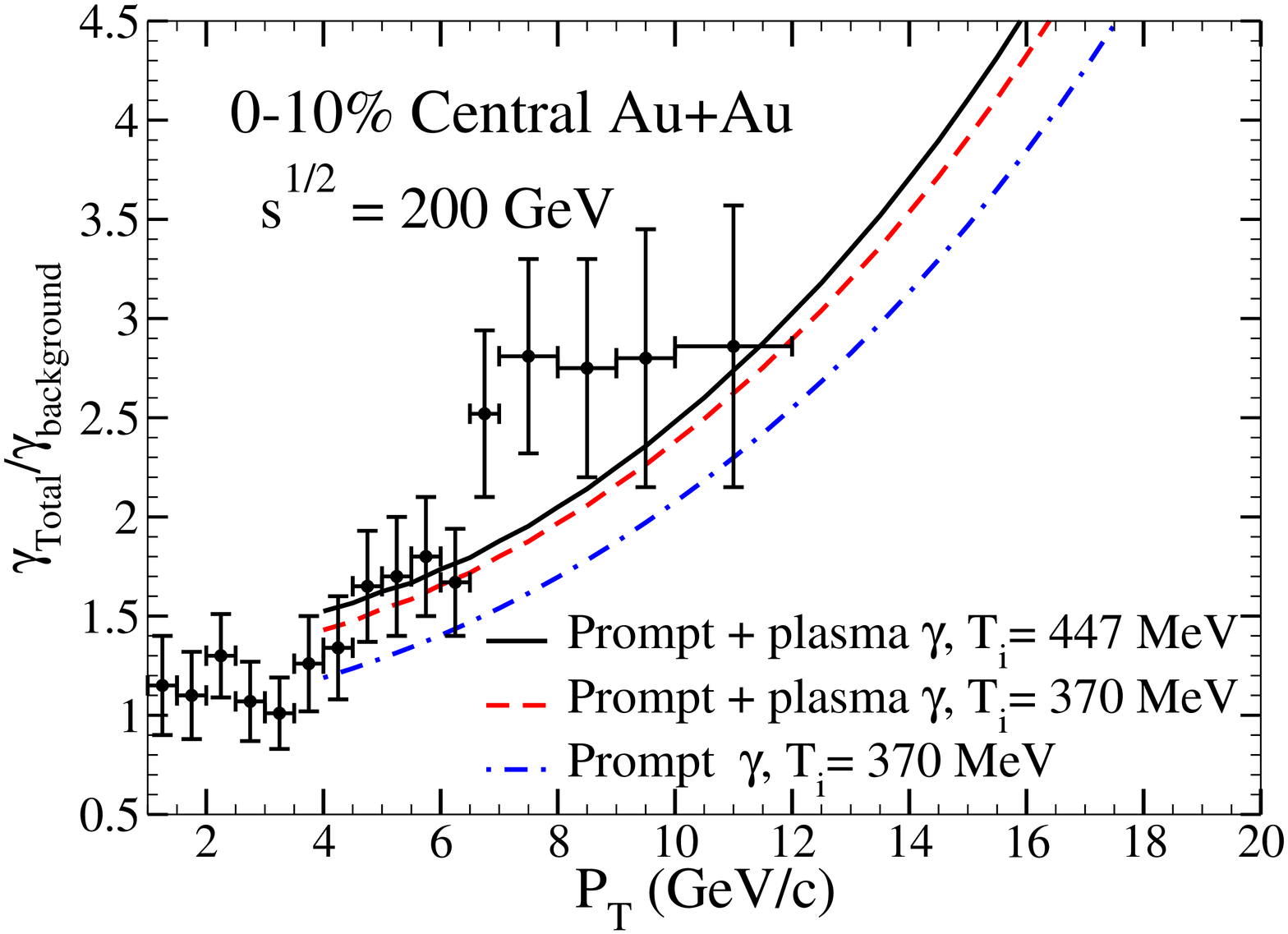,width=6.cm,angle=-90}
\psfig{file=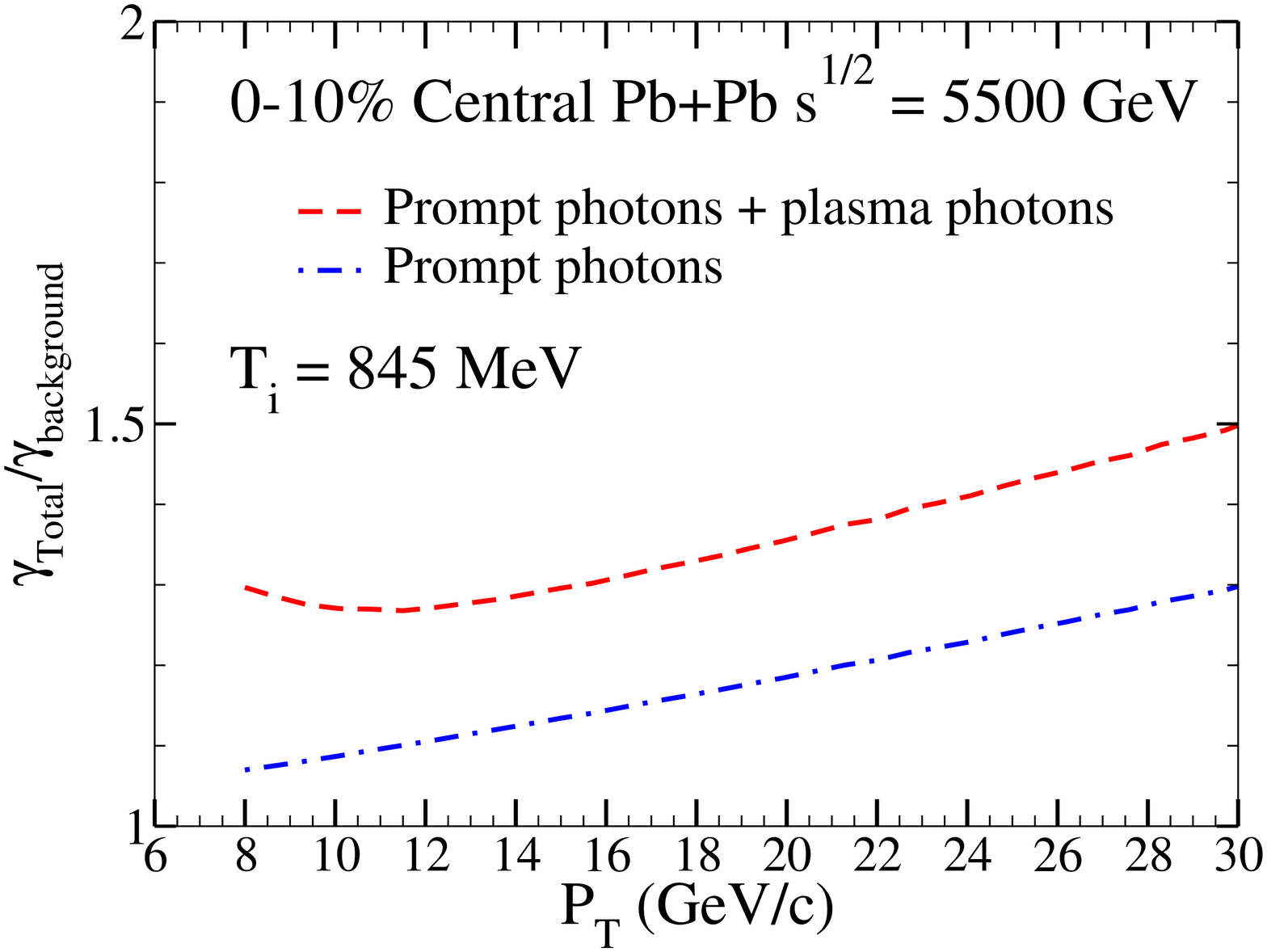,width=6.cm,angle=-90}
\caption{(Color online) Ratio of all photons over all decay photons as a
function
of $p_T$, in Au-Au collisions at RHIC (left panel) and Pb-Pb at the LHC
(right panel), with and without the thermal contribution.  In the left
panel, the effect of the initial
temperature is also shown.  The data are from the PHENIX~\cite{frantz}
collaboration.}
\label{gam_gam}
\end{figure}

Finally,  we have calculated the ratio of the total number of photons and
the background photons
\be
\label{gam_def}
\gamma_{\rm Total}/\gamma_{\rm BG}=
\frac{\frac{dN_{\gamma-\rm BG}}{d^2p_Tdy}
+\sum\mbox{all other sources}}{\frac{dN_{\gamma-\rm BG}}{d^2p_Tdy}}
\ee
and compared in Fig.~\ref{gam_gam}, with the result from
PHENIX~\cite{frantz}, with and without the QGP contribution.  The
calculation including the QGP contribution is in agreement with the data
from RHIC, except for a few data points in the range
$7 < p_T < 9$ GeV.  Without the thermal contributions, the resulting line
(dot-dashed) does not overlap at all with the experimental data.  That
could constitute a signature of the importance of the jet-photon
conversion inside the QGP, since this is the most important thermal
process as we have seen in Fig.~\ref{rhic_source}.  We also show
the weak effect of the initial temperature.  The ratio
$\gamma_{\rm Total}/\gamma_{\rm BG}$ at $T_i=447$ MeV is only enhanced by $\sim
5 \%$ relatively to the result at $T_i$=370 MeV.
Finally at the LHC (right panel), the thermal contribution is also
visible: including the
photons from the thermal phase enhances the calculation by $\sim 15\%$.

\section{Summary and Conclusions}

We have used a complete leading-order treatment of jet
energy loss in the QCD plasma to
calculate the pion and photon spectra for both RHIC and the LHC.  The
calculations
have been confronted with available data from RHIC and turn out to be in good
agreement.   These results reinforce the idea that high $p_T$
suppression is a final state effect caused by jet energy loss through
bremsstrahlung in the hot medium.

The neutral pion nuclear modification factor at RHIC as been reproduced with an
initial temperature $T_i=$370 MeV and a formation time $\tau_i$=0.26 fm/c,
corresponding to $dN/dy$=1260.  Those parameters are consistent with the
analysis in \cite{simon}. 
$R_{AA}$ has shown a large $dN/dy$
dependence, but a weak dependence on the initial temperature $T_i$
(provided the starting time $\tau_i$ is changed to keep $dN/dy$ constant).
The calculation included the nuclear geometry; assuming that all jets
are produced at the center overestimates the suppression by
$\sim 50\%$.

We have also computed the production of high $p_T$ photons from the
initial collision, from the medium, and from jet-medium interactions.
The jet-medium photons improve the agreement between experiment and
theory at RHIC, and they are expected to dominate the signal at the LHC
below about 14 GeV.  Thermal photons from the medium are not very
important to either experiment, in the kinematical range on which we have
concentrated. 
In light of these results, the in-medium production of dileptons should
also be reconsidered.  Work on this topic, and others, is in progress.

\begin{acknowledgments}
We thank P. Aurenche, R.J. Fries, J.S. Gagnon, B. M\"uller and D.K. Srivastava for useful
communications.  This work was supported in part by
the Natural Sciences and Engineering Research Council of Canada, and in part, for S.T., C.G. and S.J.,
by le Fonds Nature et Technologies du Qu\'ebec.  S.J.~also
thanks RIKEN BNL Center and U.S. Department of Energy
[DE-AC02-98CH10886] for
providing facilities essential for the completion of this work.

\end{acknowledgments}

%%%%%%%%%%%%%%%%%%%%%%%%%%%%%%%%
%%%%%%%%%%%%%%%%%%%%%%%%%%%%%%%%
%%%%%%%%%%%%%%%%%%%%%%%%%%%%%%%%
%%%%%%%%%%%%%%%%%%%%%%%%%%%%%%%%

\end{document}